\documentclass[journal]{IEEEtran}
\usepackage{amsfonts}
\usepackage{amsfonts}
\usepackage[latin1]{inputenc}
\usepackage[usenames]{color}
\usepackage[dvips]{pstcol}
\usepackage{amssymb}
\usepackage{latexsym}
\usepackage{mathrsfs}
\usepackage{amsfonts}
\usepackage{amsbsy}
\usepackage{amsmath}
\usepackage{times}
\usepackage{epsfig,epsf,times,amssymb,amsmath}
\begin{document}
\title{Adaptive Reduced-Rank Constrained Constant Modulus Beamforming Algorithms
Based on Joint Iterative Optimization of Filters  }

\author{Lei~Wang,
        ~Rodrigo~C.~de~Lamare,~and~Masahiro Yukawa
\thanks{Lei Wang and Rodrigo C. de Lamare are with the Department
of Electronics, University of York, York, YO10 5DD, U.K. (e-mail:
lw517@york.ac.uk, rcdl500@ohm.york.ac.uk).}\\
\thanks{Masahiro Yukawa is with Laboratory for Mathematical Neuroscience, BSI, RIKEN,
JAPAN (e-mail:myukawa@riken.jp).}}

\maketitle

\begin{abstract}
This paper proposes a robust reduced-rank scheme for adaptive
beamforming based on joint iterative optimization (JIO) of adaptive
filters. The novel scheme is designed according to the constant
modulus (CM) criterion subject to different constraints, and
consists of a bank of full-rank adaptive filters that forms the
transformation matrix, and an adaptive reduced-rank filter that
operates at the output of the bank of filters to estimate the
desired signal. We describe the proposed scheme for both the
direct-form processor (DFP) and the generalized sidelobe canceller
(GSC) structures. For each structure, we derive stochastic gradient
(SG) and recursive least squares (RLS) algorithms for its adaptive
implementation. The Gram-Schmidt (GS) technique is applied to the
adaptive algorithms for reformulating the transformation matrix and
improving performance. An automatic rank selection technique is
developed and employed to determine the most adequate rank for the
derived algorithms. The complexity and convexity analyses are
carried out. Simulation results show that the proposed algorithms
outperform the existing full-rank and reduced-rank methods in
convergence and tracking performance.

%This paper proposes a reduced-rank scheme for adaptive beamforming based on the constrained joint optimization filters. We
%employ this scheme to devise two novel reduced-rank adaptive algorithms according to the constant modulus (CM) criterion with
%different constraints. The first devised algorithm is formulated as a constrained joint optimization of a projection matrix and
%a reduced-rank filter with respect to the CM criterion subject to a constraint on the array response. The constrained constant
%modulus (CCM) expressions for the projection matrix and the reduced-rank weight vector are derived, and a low-complexity
%adaptive algorithm is presented to jointly estimate them for implementation. The second proposed algorithm is extended from the
%first one and implemented according to the CM criterion subject to a constraint on the array response and an orthogonal
%constraint on the projection matrix. The Gram-Schmidt (GS) technique is employed to achieve this orthogonal constraint and
%improve the performance. Simulation results are given to show superior performance of the proposed algorithms in comparison with
%existing methods.

%\textit{Index Terms}--Beamforming techniques, antenna array, constrained constant modulus, reduced-rank.

\end{abstract}

\section{Introduction}
Adaptive beamforming techniques have been developed to improve the
reception of a desired signal while suppressing interference at the
output of a sensor array. It is an ubiquitous task in array signal
processing with applications in radar, sonar, astronomy, and more
recently, in wireless communications
\cite{Anderson}-\cite{Vorobyov}. A number of adaptive algorithms for
the beamformer design are available and have been extensively
studied \cite{Trees}, \cite{Jian}. The most common are the linearly
constrained adaptive algorithms \cite{Frost}-\cite{Zou}. In general,
the linear constraints correspond to prior knowledge of certain
parameters such as the direction of arrival (DOA) of the desired
signal.

An important issue that is considered in adaptive beamforming is the
design criterion. Among many adaptive algorithms found in the
literature, the most promising criteria employed are the constrained
minimum variance (CMV) \cite{Trees} and the constrained constant
modulus (CCM) \cite{Jian} due to their simplicity and effectiveness.
%They optimize the minimum variance (MV) and constant modulus (CM)
%criteria, respectively, subject to the constraint defined by the DOA
%of the desired signal.
The CMV criterion aims to minimize the beamformer output power while
maintaining the array response on the DOA of the desired signal. The
CCM criterion is a positive measure \cite{Jian} of the deviation of
the beamformer output from a constant modulus condition subject to a
constraint on the array response of the desired signal. By measuring
the deviation, the CCM criterion provides more information than the
CMV for the parameter estimation of constant modulus constellations
in the beamformer design.

%Numerous adaptive algorithms have been proposed according to the
%constrained version criteria to realize the beamformer design
%\cite{Trees}, \cite{Frost}-\cite{Zou}. A popular approach is to
%deploy stochastic gradient (SG) approaches \cite{Haykin} since they
%lead to a low complexity solution for implementation although the
%convergence depends on the eigenvalue spread of the covariance
%matrix of the received vector. Conversely, the recursive
%least-squares (RLS) algorithms \cite{Haykin}-\cite{Chen2} have fast
%convergence, are insensitive to the eigenvalue spread of the input
%covariance matrix as compared with the SG algorithms but suffer from
%numerical instability and high complexity. A challenging problem
%experienced by previously reported algorithms, is that when the
%number of elements in the filter is large, the algorithms require a
%large number of samples to reach the steady-state. Furthermore, in
%dynamic scenarios, filters with many elements usually fail or
%provide poor performance in tracking signals embedded in
%interference and noise.

Numerous adaptive algorithms have been proposed according to the
constrained version criteria to realize the beamformer design
\cite{Trees}, { \cite{Frost}-\cite{Pezeshki}}. The major drawback of
the full-rank methods, such as stochastic gradient (SG)
\cite{Pezeshki2} and recursive least-squares (RLS) \cite{Via},
\cite{Haykin}, is that these methods require a large amount of
samples to reach the steady-state when the number of elements in the
filter is large. Furthermore, in dynamic scenarios, filters with
many elements usually fail or provide poor performance in tracking
signals embedded in interference and noise. Reduced-rank signal
processing was originally motivated to provide a way out of
this dilemma { \cite{Haimovich}-\cite{Scharf}}. %These techniques can improve convergence and tracking performance
%when dealing with large number of filter elements and show
%effectiveness in low sample support situations.
For the application of beamforming, reduced-rank schemes project the
received vector onto a lower dimensional subspace and perform the
filter optimization within this subspace. One of the popular
reduced-rank schemes is the multistage Wiener filter (MSWF), which
employs the minimum mean squared error (MMSE) \cite{Goldstein3}, and
its extended versions that utilize the CMV and CCM criteria were
reported in \cite{Honig}, \cite{Lamare}. Another technique that
resembles the MSWF \cite{Chen}, \cite{Burykh} is the
auxiliary-vector filtering (AVF)
\cite{Pados}, \cite{Pados2}. %The equivalence of the MSWF and AVF
%algorithms, which generate the same Krylov subspace, has been proved
%in \cite{Chen}, \cite{Burykh}.
A joint iterative optimization (JIO) scheme, which was presented
recently in \cite{Lamare2}, employs the CMV criterion with a
relative low-complexity adaptive implementation to achieve better
performance than the existing methods.

In this paper, we introduce a robust reduced-rank scheme based on
joint iterative optimization of filters with the CCM criterion in
detail and compare it with that of the CMV to show its improved
performance in the studied scenarios. The developed CCM reduced-rank
scheme consists of a bank of full-rank adaptive filters, which
constitutes the transformation matrix, and an adaptive reduced-rank
filter that operates at the output of the bank of full-rank filters.
The transformation matrix maps the received signal into a lower
dimension, which is then processed by the reduced-rank filter to
estimate the transmitted signal.  The proposed scheme provides an
iterative exchange of information between the transformation matrix
and the reduced-rank filter and thus leads to improved convergence
and tracking performance.

This paper makes two contributions:
\begin{itemize}
\item A reduced-rank scheme according to the constant modulus (CM) criterion subject to different constraints is
proposed based on the JIO of adaptive filters. This robust
reduced-rank scheme is investigated for both direct-form processor
(DFP) and the generalized sidelobe canceller (GSC) \cite{Haykin}
structures. For each structure, a family of computationally
efficient reduced-rank SG and RLS type algorithms are derived for
the proposed scheme. The Gram-Schmidt (GS) technique is employed in
the proposed algorithms to reformulate the transformation matrix for
further improving performance. An automatic rank selection technique
is developed to determine the most adequate rank for the proposed
algorithms.

\item The complexity comparison is presented to show the computational
costs of the proposed reduced-rank algorithms. An analysis of the
convergence properties for the proposed reduced-rank scheme is
carried out. Simulations are performed to show the improved
convergence and tracking performance of the proposed algorithms over
existing methods. The effectiveness of the GS and automatic rank
selection techniques for the proposed algorithms is visible in the
results.
\end{itemize}
% The developed CCM reduced-rank
%scheme is investigated for both direct-form processor (DFP) and
%generalized sidelobe canceller (GSC) \cite{Haykin} structures. For
%each structure, we propose a family of computationally efficient
%adaptive reduced-rank SG and RLS type algorithms for implementation.
%An automatic rank selection technique is considered to determine the
%most adequate rank for the proposed algorithms. The complexity
%comparison is presented and convergence analysis is carried out.
%Simulation results are performed to show the improved convergence
%and tracking performance of the proposed algorithms over the
%existing methods.
\vspace{1em}

The remainder of this paper is organized as follows: we outline a
system model for beamforming in Section II. {  Based on this model,
the full-rank and reduced-rank CCM beamformer designs are reviewed}.
The proposed reduced-rank scheme based on the CM criterion subject
to different constraints is presented in Section III, and the
proposed adaptive algorithms are detailed for implementation in
Section IV. The complexity and convergence analysis of the proposed
algorithms is carried out in Section V. Simulation results are
provided and discussed in Section VI, and conclusions are drawn in
Section VII.

%For convenience, some abbreviations are listed below:
%\begin{itemize}
%\item DOA~~~~~direction of arrival
%\item CMV~~~~~constrained minimum variance
%\item CCM~~~~~constrained constant modulus
%\item SG~~~~~~~~stochastic gradient
%\item RLS~~~~~recursive least squares
%\item PC~~~~~~~~principle component
%\item CS~~~~~~~~cross-spectral
%\item MSWF~~~~multistage Wiener filter
%\item MMSE~~~~minimum mean squared error
%\item AVF~~~~~auxiliary-vector filtering
%\item JIO~~~~~joint iterative optimization
%\item DFP~~~~~direct-form processor
%\item GSC~~~~~generalized sidelobe canceller
%\item GS~~~~~~~~Gram-Schmidt
%\end{itemize}

\section{System Model and CCM Beamformer Design}
In this section, we first describe a system model to express the
received data vector. Based on this model, the full-rank beamformer
design according to the CM criterion subject to the constraint on
the array response is introduced for the DFP and the GSC structures.

\subsection{System Model}
Let us suppose that $q$ narrowband signals impinge on a uniform
linear array (ULA) of $m$ ($m\geq q$) sensor elements. The sources
are assumed to be in the far field with DOAs
$\theta_{0}$,\ldots,$\theta_{q-1}$. The received vector $\boldsymbol
x(i)\in\mathbb C^{m\times 1}$ at the $i$th snapshot can be modeled
as
\begin{equation} \label{1}
\centering {\boldsymbol x}(i)={\boldsymbol {A}}({\boldsymbol
{\theta}}){\boldsymbol s}(i)+{\boldsymbol n}(i),~~~ i=1,\ldots,N,
\end{equation}
where $\boldsymbol{\theta}=[\theta_{0},\ldots,\theta_{q-1}]^{T}\in{
\mathbb{R}}^{q \times 1}$ is the signal DOAs, ${\boldsymbol
A}({\boldsymbol {\theta}})=[{\boldsymbol
a}(\theta_{0}),\ldots,{\boldsymbol a}(\theta_{q-1})]\in\mathbb{C}^{m
\times q}$ comprises the normalized signal steering vectors
${\boldsymbol a}(\theta_{k})=[1,e^{-2\pi
j\frac{u}{\lambda_{\textrm{c}}}cos{\theta_{k}}},\ldots$,\\$e^{-2\pi
j(m-1)\frac{u}{\lambda_{\textrm{c}}}cos{\theta_{k}}}]^{T}\in\mathbb{C}^{m
\times 1}, (k=0,\ldots,q-1)$, where $\lambda_{\textrm{c}}$ is the
wavelength and $u$ ($u=\lambda_{\textrm{c}}/2$ in general) is the
inter-element distance of the ULA, and to avoid mathematical
ambiguities, the steering vectors $\boldsymbol a(\theta_{k})$ are
assumed to be linearly independent. ${\boldsymbol s}(i)\in
\mathbb{C}^{q\times 1}$ is the source data, ${\boldsymbol
n}(i)\in\mathbb{C}^{m\times 1}$ is temporary white sensor noise,
which is assumed to be a zero-mean spatially and Gaussian process,
$N$ is the observation size of snapshots, and $(\cdot)^{T}$\ stands
for transpose. %The output of a narrowband beamformer is given by
%\begin{equation} \label{2}
%\centering y(i)={\boldsymbol w}^H(i) {\boldsymbol x}(i)
%\end{equation}
%where ${\boldsymbol
%w}(i)=[w_{1}(i),\ldots,w_{m}(i)]^{T}\in\mathbb{C}^{m\times 1}$ is
%the complex weight vector, and $(\cdot)^{H}$ stands for Hermitian
%transpose.

\subsection{  Full-rank CCM Beamformer Design}
The full-rank CCM linear receiver design for beamforming is
equivalent to determining a filter ${\boldsymbol
w}(i)=[w_{1}(i),\ldots,w_{m}(i)]^{T}\in\mathbb{C}^{m\times 1}$ that
provides an estimate of the desired symbol $y(i)=\boldsymbol
w^H(i)\boldsymbol x(i)$, where $(\cdot)^{H}$ denotes Hermitian
transpose. The calculation of the weight vector is based on the minimization of the following cost function:%The weight vector $\boldsymbol w(i)$ is calculated by
%minimizing the CM cost function
\begin{equation}\label{2}
\begin{split}
&J_{\textrm{cm}}\Big(\boldsymbol w(i)\Big)=\mathbb E\Big\{\big[{ |y(i)|^{p}}-\nu\big]^{2}\Big\},\\
&\textrm{subject~to}~~{\boldsymbol w}^{H}(i){\boldsymbol
a}(\theta_{0})=\gamma,
%&{\boldsymbol w}_{\textrm{opt}}=\arg\min_{\boldsymbol w} E\big\{\big[|y(i)|^p-R_{\textrm p}\big]^2\big\},~i=1,\ldots,N\\
%&\textrm{subject~to}~~{\boldsymbol w}^{H}(i){\boldsymbol a}(\theta_{0})=1.
\end{split}
\end{equation}
where $\nu$ is suitably chosen to guarantee that the weight solution
is close to the global minimum and $\gamma$ is set to ensure the
convexity of (\ref{2}) \cite{Lamare}. The quantity $\theta_0$ is the
direction of the desired signal, $\boldsymbol a(\theta_{0})$ denotes
the corresponding normalized steering vector, and in general, $p=2$
is selected to consider the cost function as the expected deviation
of the squared modulus of the beamformer output to a constant, say
$\nu=1$. {  The CCM criterion is a positive measure \cite{Jian} of
the deviation of the beamformer output from a constant modulus
condition subject to a constraint on the array response of the
desired signal. Compared with the CMV criterion, it exploits a
constant modulus property of the transmitted signals, utilizes the
deviation to provide more information for the parameter estimation
of the constant modulus constellations, and achieves a superior
performance \cite{Lamare3}, \cite{Lamare}}. The CCM beamformer
minimizes the contribution of interference and noise while
maintaining the gain along the look direction to be constant. The
weight expression of the full-rank CCM design is given in
\cite{Lamare}.

\subsection{  Reduced-rank CCM Beamformer Design}
\begin{figure}[htb]
\begin{minipage}[h]{1.0\linewidth}
  \centering
  \centerline{\epsfig{figure=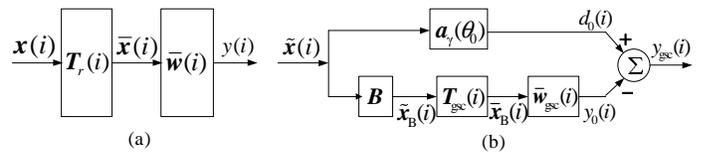,scale=0.68}} \vspace{-0em}\caption{Reduced-rank scheme for (a) the DFP and (b) the GSC structures.} \label{fig:model22}
\end{minipage}
\end{figure}

For large $m$, considering the high computational cost and poor
performance associated with the full-rank filter, a number of recent
works in the literature have been reported based on reduced-rank
schemes { \cite{Haimovich}-\cite{Goldstein2}},
\cite{Goldstein3}-\cite{Wong}. Here, we will describe a reduced-rank
framework that reduces the number of coefficients by mapping the
received vector into a lower dimensional subspace. {  The diagrams
of the reduced-rank processors are depicted for the DFP and the GSC
structures in Fig. \ref{fig:model22}(a) and Fig.
\ref{fig:model22}(b), respectively.}

\subsubsection{  Beamformer Design for the DFP}
In the DFP structure, $\boldsymbol T_r\in\mathbb C^{m\times r}$
denotes the transformation matrix that includes a set of $m\times 1$
vectors for a $r$-dimensional subspace with $r\leq m$. The
transformation matrix maps the received vector $\boldsymbol x(i)$
into its low-dimension version $\bar{\boldsymbol x}(i)\in\mathbb
C^{r\times 1}$, which is given by
\begin{equation}\label{3}
\bar{\boldsymbol x}(i)=\boldsymbol T_r^H(i)\boldsymbol x(i),
\end{equation}
where, in what follows, all $r$-dimensional quantities are denoted
by an over bar. An adaptive reduced-rank CCM filter
$\bar{\boldsymbol w}(i)\in\mathbb C^{r\times 1}$ follows the
transformation matrix to produce the filter output
$y(i)=\bar{\boldsymbol w}^H(i)\bar{\boldsymbol x}(i)$.

{  Substituting the expression of $y(i)$ into the cost function in
(\ref{2}) and calculating for the reduced-rank weight vector, we
have \cite{Lamare}}
\begin{equation}\label{4}
\bar{\boldsymbol w}(i+1)=\bar{\boldsymbol
R}^{-1}(i)\Big\{\bar{\boldsymbol p}(i)-\frac{\big[\bar{\boldsymbol
p}^H(i)\bar{\boldsymbol R}^{-1}(i)\bar{\boldsymbol
a}(\theta_0)-\gamma\big]\bar{\boldsymbol
a}(\theta_0)}{\bar{\boldsymbol a}^H(\theta_0)\bar{\boldsymbol
R}^{-1}(i)\bar{\boldsymbol a}(\theta_0)}\Big\},
\end{equation}
where $\bar{\boldsymbol R}(i)=\mathbb E[|y(i)|^2\boldsymbol
T_r^H(i)\boldsymbol x(i)\boldsymbol x^H(i)\boldsymbol
T_r(i)]\in\mathbb C^{r\times r}$, $\bar{\boldsymbol
a}(\theta_0)=\boldsymbol T_r^H\boldsymbol a(\theta_0)\in\mathbb
C^{r\times 1}$, and $\bar{\boldsymbol p}(i)=\mathbb
E[y^{\ast}(i)\boldsymbol T_r^H(i)\boldsymbol x(i)]\in\mathbb
C^{r\times 1}$. {  Note that the expression in (\ref{4}) is a
function of previous values of $\bar{\boldsymbol w}(i)$ (since
$y(i)=\bar{\boldsymbol w}^H(i)\bar{\boldsymbol x}(i)$) and thus must
be initialized to start the computation for the solution. We keep
the time index in $\bar{\boldsymbol R}(i)$ and $\bar{\boldsymbol
p}(i)$ for the same reason.}

\subsubsection{  Beamformer Design for the GSC}
The GSC structure converts the constrained optimization problem into
an unconstrained one and adopts an alternative way to realize the
beamformer design. {  The full-rank CCM beamformer design with
respect to the GSC structure has been reported in \cite{Rude}. Here,
we employ an alternative way proposed in \cite{Chern2}, \cite{Chern}
to describe a reduced-rank GSC structure.} As can be seen in Fig.
\ref{fig:model22}(b), the reduced-rank GSC structure composes a
constrained component { ($\boldsymbol
a_{\gamma}(\theta_0)=\gamma\boldsymbol a(\theta_0)$) and an
unconstrained component.} $\tilde{\boldsymbol x}(i)$ is a new
received vector defined as
\begin{equation}\label{5}
\tilde{\boldsymbol x}(i)=y_{\textrm{gsc}}^{\ast}(i)\boldsymbol x(i),
\end{equation}
where $y_{\textrm{gsc}}(i)=\boldsymbol w^H(i)\boldsymbol x(i)$. {
The definition of $\tilde{\boldsymbol x}(i)$ is valid for $p=2$ in
(\ref{2}) and $|y_{\textrm{gsc}}(i)|^2=\boldsymbol
w^H(i)\tilde{\boldsymbol x}(i)$. This expression is only to favor
its use in the GSC structure for the case of the CM cost function.
Note that $y_{\textrm{gsc}}(i)$ and $y(i)$ (full-rank or
reduced-rank with $\boldsymbol T_r=\boldsymbol I_{m\times m}$)
correspond to the same values but are written in a different way to
indicate the structures (DFP and GSC).}

{  For the constrained component}, the output is $d_0(i)=\boldsymbol
a_{\gamma}^H(\theta_0)\tilde{\boldsymbol x}(i)$. {  With respect to
the unconstrained component, the new received vector passes through
a signal blocking matrix $\boldsymbol B\in\mathbb C^{(m-1)\times m}$
to get a transformed vector $\tilde{\boldsymbol x}_B(i)\in\mathbb
C^{(m-1)\times1}$}, which is
\begin{equation}\label{6}
\tilde{\boldsymbol x}_{B}(i)=\boldsymbol B\tilde{\boldsymbol x}(i),
\end{equation}
where $\boldsymbol B$ is obtained by the singular value
decomposition or the QR decomposition algorithms \cite{Goldstein4}.
Thus, $\boldsymbol B\boldsymbol a(\theta_0)=\boldsymbol
0_{(m-1)\times 1}$ means that the term $\boldsymbol B$ effectively
blocks any signal coming from the look direction $\theta_0$. The
transformation matrix $\boldsymbol T_{\textrm{gsc}}(i)\in\mathbb
C^{(m-1)\times r}$ maps the transformed vector $\tilde{\boldsymbol
x}_B(i)$ into a low-dimension version, as described by
\begin{equation}\label{7}
\bar{\boldsymbol x}_B(i)=\boldsymbol
T_{\textrm{gsc}}^H(i)\tilde{\boldsymbol x}_B(i).
\end{equation}

The reduced-rank received vector $\bar{\boldsymbol x}_B(i)$ is
processed by a reduced-rank filter $\bar{\boldsymbol
w}_{\textrm{gsc}}(i)\in\mathbb C^{r\times1}$ to get the
unconstrained output $y_0(i)=\bar{\boldsymbol
w}_{\textrm{gsc}}^H(i)\bar{\boldsymbol x}_B(i)$. The reduced-rank
weight vector is \cite{Haykin}
\begin{equation}\label{8}
\bar{\boldsymbol w}_{\textrm{gsc}}(i+1)=\bar{\boldsymbol
R}_{\bar{x}_B}^{-1}(i)\bar{\boldsymbol p}_{B}(i),
\end{equation}
where $\bar{\boldsymbol R}_{\bar{x}_B}(i)=\mathbb E[\boldsymbol
T_{\textrm{gsc}}^H(i)\tilde{\boldsymbol x}_{B}(i)\tilde{\boldsymbol
x}_{B}^H(i)\boldsymbol T_{\textrm{gsc}}(i)]\in\mathbb C^{r\times r}$
and $\bar{\boldsymbol p}_{B}(i)=\mathbb
E[(d_0^{\ast}(i)-1)\boldsymbol
T_{\textrm{gsc}}^H(i)\tilde{\boldsymbol x}_{B}(i)]\in\mathbb
C^{r\times 1}$. {  Note that this expression is a function of
previous values of the weight vector and therefore must be
initialized to start the computation for the solution.}

The reduced-rank GSC structure can be concluded in a transformation
operator $\bar{\boldsymbol S}=[\boldsymbol a_{\gamma}(\theta_0),
\boldsymbol B^H\boldsymbol T_{\textrm{gsc}}]^H\in\mathbb
C^{(r+1)}\times m$ and a reduced-rank weight vector
$\bar{\boldsymbol w}'=[1,-\bar{\boldsymbol
w}_{\textrm{gsc}}^H]^H\in\mathbb C^{(r+1)\times1}$. The equivalent
full-rank weight vector can be expressed as
\begin{equation}\label{9}
\begin{split}
\boldsymbol w(i+1)&=\bar{\boldsymbol S}^H\bar{\boldsymbol w}'(i+1)\\
&=\boldsymbol a_{\gamma}(\theta_0)-\boldsymbol B^H\boldsymbol
T_{\textrm{gsc}}(i+1)\bar{\boldsymbol w}_{\textrm{gsc}}(i+1).
\end{split}
\end{equation}

{  The reduced-rank weight expressions in (\ref{4}) for the DFP and
in (\ref{9}) for the GSC are general forms to the signal processing
tasks. Specifically, for $r=m$ (DFP) and $r=m-1$ (GSC), the
expressions are equivalent to the full-rank filtering schemes
\cite{Haykin}. For $1<r<m$ (DFP) and $1<r<m-1$ (GSC), the signal
processing tasks are changed and the reduced-rank filters estimate
the desired signals.}

The challenge left to us is how to efficiently design and calculate
the transformation matrices $\boldsymbol T_r$ and $\boldsymbol
T_{\textrm{gsc}}$. The {  principal components} (PC) method reported
in \cite{Haimovich} uses the eigenvectors of the interference-only
covariance matrix corresponding to the eigenvalues of significant
magnitude to construct the transformation matrix. The {
cross-spectral} (CS) method \cite{Goldstein}, a counterpart of the
PC method belonging to the eigen-decomposition family, forms the
transformation matrix by using the eigenvectors which contribute the
most towards maximizing the SINR and outperforms the PC method.
Another family of adaptive reduced-rank filters such as the MSWF
\cite{Goldstein3}, \cite{Honig} and the AVF \cite{Pados} generates a
set of basis vectors as the transformation matrix that spans the
same Krylov subspace \cite{Chen}, \cite{Burykh}.

\section{Proposed CCM reduced-rank scheme}

In this section, we introduce the proposed reduced-rank scheme based
on the JIO approach. Two optimization problems according to the CM
criterion subject to different constraints are described for the
proposed scheme. Based on this scheme, we derive the expressions of
the transformation matrix and the reduced-rank weight vector. For
the sake of completeness, the proposed scheme is realized for both
the DFP and the GSC structures.

%In this section, we introduce the proposed CCM reduced-rank scheme
%based on the JIO approach reported in \cite{Lamare2}, which utilizes
%the CMV as the design criterion. Compared with the CMV criterion,
%the CCM criterion is a positive measure of the average amount that
%the beamformer output deviates from a constant modulus condition
%\cite{Jian}, which provides more information for the parameter
%estimation in the beamformer design. Based on the proposed
%reduced-rank scheme, we derive the expressions of the projection
%matrix and the reduced-rank weight vector. The proposed scheme is
%realized for both the DFP and GSC structures. %Under each structure, we
%employ SG and RLS type algorithms for adaptive implementations of
%the proposed scheme. An automatic rank selection technique is
%described to determine the most adequate rank for the proposed
%algorithms.

\subsection{Proposed CCM Reduced-rank Scheme for the DFP}

\begin{figure}[htb]
\begin{minipage}[h]{1.0\linewidth}
  \centering
  \centerline{\epsfig{figure=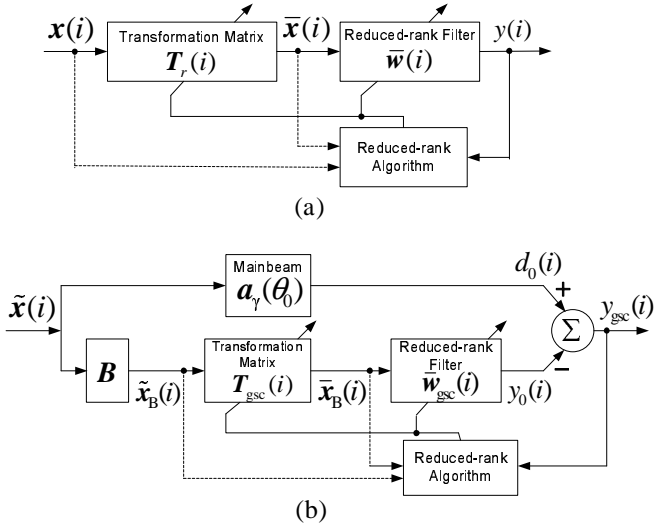,scale=0.65}} \vspace{-0em}\caption{Proposed reduced-rank scheme for (a) the DFP and (b) the GSC structures.} \label{fig:model33}
\end{minipage}
\end{figure}

Here we detail the principles of the proposed CCM reduced-rank
scheme using a transformation based on adaptive filters. For the DFP
structure depicted in Fig. \ref{fig:model33}(a), the proposed scheme
employs a transformation matrix $\boldsymbol T_r(i)\in\mathbb
C^{m\times r}$, which is responsible for the dimensionality
reduction, to generate $\bar{\boldsymbol x}(i)\in\mathbb C^{r\times
1}$. The dimension is reduced and the key features of the original
signal is retained in $\bar{\boldsymbol x}(i)$ according to the CCM
criterion. The transformation matrix is structured as a bank of $r$
full-rank filters $\boldsymbol t_j(i)=[t_{1,j}(i), t_{2,j}(i),
\ldots, t_{m,j}(i)]^T\in\mathbb C^{m\times 1}$, $(j=1, \ldots, r)$
as given by $\boldsymbol T_r(i)=[\boldsymbol t_1(i), \boldsymbol
t_2(i), \ldots, \boldsymbol t_r(i)]$. An adaptive reduced-rank
filter $\bar{\boldsymbol w}(i)\in\mathbb C^{r\times 1}$ is then used
to produce the output. The transformation matrix $\boldsymbol
T_r(i)$ and the reduced-rank filter $\bar{\boldsymbol w}(i)$ are
jointly optimized in the proposed scheme. The filter output is a
function of the received vector, the transformation matrix, and the
reduced-rank weight vector, which is
%\begin{figure}[!htb]
%\begin{center}
%\def\epsfsize#1#2{1.0\columnwidth}
%\epsfbox{model33.eps} \caption{system model33.} \label{fig:model33}
%\end{center}
%\end{figure}
\begin{equation}\label{10}
y(i)=\bar{\boldsymbol w}^H(i)\boldsymbol T_r^H(i)\boldsymbol
x(i)=\bar{\boldsymbol w}^H(i)\bar{\boldsymbol x}(i).
\end{equation}

We describe two optimization problems according to the CM cost
function subject to different constraints for the proposed
reduced-rank scheme, which are given by
\begin{equation}\label{11}
\begin{split}
\textrm{Problem 1}:~&\min~J_{\textrm{cm}}\big(\boldsymbol T_r(i),
\bar{\boldsymbol
w}(i)\big)=\mathbb E\Big\{\big[|y(i)|^2-1\big]^2\Big\}\\
&\textrm{subject~to}~\bar{\boldsymbol w}^H(i)\boldsymbol
T_r^H(i)\boldsymbol a(\theta_0)=\gamma,
\end{split}
\end{equation}
\begin{equation}\label{12}
\begin{split}
&\textrm{Problem 2}:~\min~J_{\textrm{cm}}\big(\boldsymbol T_r(i),
\bar{\boldsymbol
w}(i)\big)=\mathbb E\Big\{\big[|y(i)|^2-1\big]^2\Big\}\\
&\textrm{subject~to}~\bar{\boldsymbol w}^H(i)\boldsymbol
T_r^H(i)\boldsymbol a(\theta_0)=\gamma~\textrm{and}~\boldsymbol
T_r^H(i)\boldsymbol T_r(i)=\boldsymbol I.
\end{split}
\end{equation}

Compared with (\ref{11}), the problem in (\ref{12}) has an
orthogonal constraint on the transformation matrix, which is to
reformulate $\boldsymbol T_r(i)$. {  The transformation matrix
generated from (\ref{11}) has vectors that may perform a similar
operation (e.g., take the same information twice or more), thereby
making poor use of the data and losing performance. The subspace
computed with (\ref{12}), which spans the same subspace as
$\boldsymbol T_r(i)$, generates basis vectors that are orthogonal to
each other and which does not affect the noise statistically. The
reformulated transformation matrix performs an efficient operation
to keep all useful information in the generated reduced-rank
received vector, which is important to estimate the desired signal
and improve the performance.} In the following, we will derive the
CCM expressions of $\boldsymbol T_r(i)$ and $\bar{\boldsymbol w}(i)$
for solving (\ref{11}) and (\ref{12}).

The cost function in (\ref{11}) can be transformed by the method of
Lagrange multipliers into an unconstrained one, which is
\begin{equation}\label{13}
\begin{split}
L_{\textrm{cm}}\big(\boldsymbol T_r(i), \bar{\boldsymbol
w}(i)\big)&=\mathbb E\Big\{\big[|\bar{\boldsymbol w}^H(i)\boldsymbol
T_r^H(i)\boldsymbol
x(i)|^2-1\big]^2\Big\}\\
&+2\mathfrak{R}\Big\{\lambda\big[\bar{\boldsymbol w}^H(i)\boldsymbol
T_r^H(i)\boldsymbol a(\theta_0)-\gamma\big]\Big\},
\end{split}
\end{equation}
where $\lambda$ is a scalar Lagrange multiplier and the operator
$\mathfrak{R}(\cdot)$ selects the real part of the argument.

Assuming $\bar{\boldsymbol w}(i)$ is known, {  computing the
gradient of (\ref{13}) with respect to $\boldsymbol T_r(i)$ (matrix
calculus), equating it to a null matrix and solving for $\lambda$},
we have
\begin{equation}\label{14}
\begin{split}
&\boldsymbol T_r(i+1)=\boldsymbol R^{-1}(i)\Big\{\boldsymbol
p(i)\bar{\boldsymbol w}^H(i)-\\
&\frac{\big[\bar{\boldsymbol w}^H(i)\bar{\boldsymbol
R}_{\bar{w}}^{-1}(i)\bar{\boldsymbol w}(i)\boldsymbol
p^H(i)\boldsymbol R^{-1}(i)\boldsymbol
a(\theta_0)-\gamma\big]\boldsymbol a(\theta_0)\bar{\boldsymbol
w}^H(i)}{\bar{\boldsymbol w}^H(i)\bar{\boldsymbol
R}_{\bar{w}}^{-1}(i)\bar{\boldsymbol w}(i)\boldsymbol
a^H(\theta_0)\boldsymbol R^{-1}(i)\boldsymbol
a(\theta_0)}\Big\}\bar{\boldsymbol R}_{\bar{w}}^{-1}(i),
\end{split}
\end{equation}
where $\boldsymbol p(i)=\mathbb E[y^{\ast}(i)\boldsymbol
x(i)]\in\mathbb C^{m\times 1}$, $\boldsymbol R(i)=\mathbb
E[|y(i)|^2\boldsymbol x(i)\boldsymbol x^H(i)]\in\mathbb C^{m\times
m}$, and $\bar{\boldsymbol R}_{\bar{w}}(i)=\mathbb
E[\bar{\boldsymbol w}(i)\bar{\boldsymbol w}^H(i)]\in\mathbb
C^{r\times r}$. Note that the reduced-rank weight vector
$\bar{\boldsymbol w}(i)$ depends on the received vectors that are
random in practice, thus $\bar{\boldsymbol R}_{\bar{w}}(i)$ is
$r$-rank and invertible. $\boldsymbol R(i)$ and $\boldsymbol p(i)$
are functions of previous values of $\boldsymbol T_r(i)$ and
$\bar{\boldsymbol w}(i)$ due to the presence of $y(i)$. Therefore,
it is necessary to initialize $\boldsymbol T_r(i)$ and
$\bar{\boldsymbol w}(i)$ to estimate $\boldsymbol R(i)$ and
$\boldsymbol p(i)$, and start the computation.

On the other hand, assuming $\boldsymbol T_r(i)$ is known, {
computing the gradient of (\ref{13}) with respect to
$\bar{\boldsymbol w}(i)$, equating it to a null vector, and solving
for $\lambda$}, we obtain
\begin{equation}\label{15}
\bar{\boldsymbol w}(i+1)=\bar{\boldsymbol
R}^{-1}(i)\Big\{\bar{\boldsymbol p}(i)-\frac{\big[\bar{\boldsymbol
p}^H(i)\bar{\boldsymbol R}^{-1}(i)\bar{\boldsymbol
a}(\theta_0)-\gamma\big]\bar{\boldsymbol
a}(\theta_0)}{\bar{\boldsymbol a}^H(\theta_0)\bar{\boldsymbol
R}^{-1}(i)\bar{\boldsymbol a}(\theta_0)}\Big\},
\end{equation}
where $\bar{\boldsymbol R}(i)=\mathbb E[|y(i)|^2\boldsymbol
T_r^H(i)\boldsymbol x(i)\boldsymbol x^H(i)\boldsymbol
T_r(i)]\in\mathbb C^{r\times r}$, $\bar{\boldsymbol p}(i)=\mathbb
E[y^{\ast}(i)\boldsymbol T_r^H(i)\boldsymbol x(i)]\in\mathbb
C^{r\times 1}$, and $\bar{\boldsymbol a}(\theta_0)=\boldsymbol
T_r^H(i)\boldsymbol a(\theta_0)$.

Note that the expressions in (\ref{14}) for the transformation
matrix and (\ref{15}) for the reduced-rank weight vector can be
applied to solve the optimization problem (\ref{12}). The orthogonal
constraint in (\ref{12}) can be realized by the Gram-Schmidt (GS)
technique, which will be illustrated in the next section.

\subsection{Proposed CCM Reduced-rank Scheme for the GSC}
For the GSC structure, as depicted in Fig. \ref{fig:model33}(b), the
proposed scheme utilizes a transformation matrix $\boldsymbol
T_{\textrm{gsc}}(i)\in\mathbb C^{(m-1)\times r}$ to map the new
transformed vector $\tilde{\boldsymbol x}_{B}(i)\in\mathbb
C^{(m-1)\times 1}$ into a lower dimension, say $\bar{\boldsymbol
x}_{B}(i)=\boldsymbol T_{\textrm{gsc}}^H(i)\tilde{\boldsymbol
x}_{B}(i)\in\mathbb C^{r\times 1}$. In our design, the
transformation matrix $\boldsymbol T_{\textrm{gsc}}(i)$ and the
reduced-rank weight vector $\bar{\boldsymbol w}_{\textrm{gsc}}(i)$
for the sidelobe of the GSC are jointly optimized by minimizing the
cost function
\begin{equation}\label{16}
\begin{split}
&J_{\textrm{cm-gsc}}\big(\boldsymbol T_{\textrm{gsc}}(i),
\bar{\boldsymbol
w}_{\textrm{gsc}}(i)\big)=\\
&~~~~~~~~~~\mathbb E\Big\{\big[\big(\boldsymbol
a_{\gamma}(\theta_0)-\boldsymbol B^H\boldsymbol
T_{\textrm{gsc}}(i)\bar{\boldsymbol
w}_{\textrm{gsc}}(i)\big)^H\tilde{\boldsymbol x}(i)-1\big]^2\Big\},
\end{split}
\end{equation}
where the expression in (\ref{16}) for the GSC is obtained by {
substituting (\ref{5}) and (\ref{9}) into (\ref{2}) with $p=2$}.
This is an unconstrained cost function that corresponds to
(\ref{11}). From Fig. \ref{fig:model33} (b), this structure
essentially decomposes the adaptive weight vector into constrained
(array response) and unconstrained components (see also Eq.
(\ref{9})). The unconstrained component can be adjusted to meet the
CM criterion since the constrained component always ensures that the
constrained condition is satisfied. Thus, the proposed GSC framework
converts the constrained optimization problem into an unconstrained
one.

Assuming $\bar{\boldsymbol w}_{\textrm{gsc}}(i)$ and $\boldsymbol
T_{\textrm{gsc}}(i)$ are given, respectively, {  computing the
gradient} of (\ref{16}) with respect to $\boldsymbol
T_{\textrm{gsc}}(i)$ and $\bar{\boldsymbol w}_{\textrm{gsc}}(i)$,
and solving the equations yields
\begin{equation}\label{17}
\boldsymbol T_{\textrm{gsc}}(i+1)=\boldsymbol
R_{\tilde{x}_{B}}^{-1}(i)\tilde{\boldsymbol
p}_{B}(i)\bar{\boldsymbol w}_{\textrm{gsc}}^H(i)\bar{\boldsymbol
R}_{\bar{w}_{\textrm{gsc}}}^{-1}(i),
\end{equation}
\begin{equation}\label{18}
\bar{\boldsymbol w}_{\textrm{gsc}}(i+1)=\bar{\boldsymbol
R}_{\bar{x}_{B}}^{-1}(i)\bar{\boldsymbol p}_{B}(i),
\end{equation}
where $\boldsymbol R_{\tilde{\boldsymbol x}_{B}}(i)$ and
$\tilde{\boldsymbol p}_{B}(i)$ have been defined in the previous
section, and $\bar{\boldsymbol
R}_{\bar{w}_{\textrm{gsc}}}(i)=\mathbb E[\bar{\boldsymbol
w}_{\textrm{gsc}}(i)\bar{\boldsymbol
w}_{\textrm{gsc}}^H(i)]\in\mathbb C^{r\times r}$. $\bar{\boldsymbol
R}_{\bar{w}_\textrm{gsc}}(i)$ is invertible since $\bar{\boldsymbol
w}_{\textrm{gsc}}(i)$ depends on the random received vector and
$\bar{\boldsymbol R}_{\bar{w}_\textrm{gsc}}(i)$ is a full-rank
matrix. $\bar{\boldsymbol R}_{\bar{x}_B}(i)=\mathbb E[\boldsymbol
T_{\textrm{gsc}}^H(i)\tilde{\boldsymbol x}_{B}(i)\tilde{\boldsymbol
x}_{B}^H(i)\boldsymbol T_{\textrm{gsc}}(i)]\in\mathbb C^{r\times r}$
and $\bar{\boldsymbol p}_{B}(i)=\mathbb
E[(d_0^{\ast}(i)-1)\boldsymbol
T_{\textrm{gsc}}^H(i)\tilde{\boldsymbol x}_{B}(i)]\in\mathbb
C^{r\times 1}$. Again, the orthogonal constraint on the
transformation matrix can be enforced in the optimization problem
(\ref{16}) and the GS technique is employed to realize this.

Note that the filter expressions in (\ref{14}) and (\ref{15}) for
the DFP and (\ref{17}) and (\ref{18}) for the GSC are not
closed-form solutions. In the DFP structure, the expression of the
transformation matrix in (\ref{14}) is a function of
$\bar{\boldsymbol w}(i)$ and the reduced-rank weight vector obtained
from (\ref{15}) depends on $\boldsymbol T_{r}(i)$. It is necessary
to set initial values of $\boldsymbol T_{r}(i)$ and
$\bar{\boldsymbol w}(i)$ for the update procedures. Thus,
initialization about the transformation matrix and the reduced-rank
weight vector is not only to get a beamformer output $y(i)$ for
estimating $\boldsymbol R(i)$ and $\bar{\boldsymbol R}(i)$, but to
start the computation of the proposed scheme. In the case of the
GSC, we initialize $\boldsymbol T_{\textrm{gsc}}(i)$ and
$\bar{\boldsymbol w}_{\textrm{gsc}}(i)$ with the same intention.

Unlike the MSWF \cite{Goldstein3} and the AVF \cite{Pados}
techniques in which the transformation matrix is computed
independently from the reduced-rank filter, the proposed scheme
provides an iterative exchange of information between the
transformation matrix and the reduced-rank filter, which leads to
improved convergence and tracking performance. The transformation
matrix reduces the dimension of the received vector whereas the
reduced-rank filter attempts to estimate the desired signal. The key
strategy lies in the joint iterative optimization of the filters. In
the next section, we will derive iterative solutions via simple
adaptive algorithms and introduce an automatic rank selection
technique for the adaptation of the rank $r$.

\section{Adaptive Algorithms of The Proposed CCM Reduced-rank Scheme}
We derive SG and RLS type algorithms for the proposed CCM
reduced-rank scheme. Some related works can be found in {
\cite{Pezeshki}-\cite{Via}}. In this paper, the adaptive algorithms
are described for the DFP and the GSC structures, respectively, to
perform joint iterative updates of the transformation matrix and the
reduced-rank weight vector. They are used to solve Problem 1. The
Gram-Schmidt (GS) technique is employed in these algorithms and
imposes an orthogonal constraint on the transformation matrix { to
solve Problem 2}. An automatic rank selection technique is
introduced to determine the most adequate rank for the proposed
methods.

\subsection{Stochastic Gradient Algorithms}
Here, we derive the SG algorithms with the proposed CCM reduced-rank
scheme for both the DFP and the GSC structures.
\subsubsection{SG algorithm for the DFP}
Assuming $\bar{\boldsymbol w}(i)$ and ${\boldsymbol T}_{r}(i)$ are
known, respectively, computing the instantaneous gradient of
(\ref{13}) with respect to $\boldsymbol T_r(i)$ and
$\bar{\boldsymbol w}(i)$, we obtain
\begin{equation}\label{19}
\nabla L_{\textrm{cm}_{T_r(i)}}(i)=2e(i)y^{\ast}(i)\boldsymbol
x(i)\bar{\boldsymbol w}^H(i)+2\lambda_{T_r}^{\ast}\boldsymbol
a(\theta_0)\bar{\boldsymbol w}^H(i),
\end{equation}
\begin{equation}\label{20}
\nabla L_{\textrm{cm}_{\bar{w}(i)}}(i)=2e(i)y^{\ast}(i)\boldsymbol
T_r^H(i)\boldsymbol x(i)+2\lambda_{\bar{w}}^{\ast}\boldsymbol
T_r^H(i)\boldsymbol a(\theta_0),
\end{equation}
where $e(i)=|y(i)|^2-1$.

Following the gradient rules $\boldsymbol T_r(i+1)=\boldsymbol
T_r(i)-\mu_{T_r}\nabla J_{\textrm{un}_{T_r(i)}}(i)$ and
$\bar{\boldsymbol w}(i+1)=\bar{\boldsymbol w}(i)-\mu_{\bar{w}}\nabla
J_{\textrm{un}_{\bar{w}(i)}}(i)$, substituting (\ref{19}) and
(\ref{20}) into them, respectively, and solving the Lagrange
multipliers $\lambda_{T_r}$ and $\lambda_{\bar{w}}$ by employing the
constraint in (\ref{11}), we obtain the iterative SG algorithm for
the DFP, {  which is denominated as JIO-CCM-SG. A summary of this
algorithm is given in Table \ref{tab:JIO-CCM-SG-DFP}}, where
$\mu_{T_r}$ and $\mu_{\bar{w}}$ are the corresponding step size
factors for the DFP, which are small positive values. The
initialization values are set to satisfy the constraint in
(\ref{11}). The transformation matrix $\boldsymbol T_r(i)$ and the
reduced-rank weight vector $\bar{\boldsymbol w}(i)$ operate together
and exchange information at each time instant.
\begin{table}[!t]
\centering
    \caption{THE JIO-CCM-SG ALGORITHM FOR DFP}     % NOTE!  caption goes _before_ the table contents !!
    \label{tab:JIO-CCM-SG-DFP}
    \begin{small}
        \begin{tabular}{l}
\hline
\bfseries {Initialization:}\\
${\boldsymbol T}_r(1)=[{\boldsymbol I}_{r\times r}~\boldsymbol 0_{r\times (m-r)}]^T$;\\
${\bar{\boldsymbol w}}(1)=\boldsymbol T_r^H(1)\boldsymbol a_{\gamma}(\theta_0)/(\|\boldsymbol T_r^H(1)\boldsymbol a_{\gamma}(\theta_0)\|^2)$.\\
\bfseries {Update for each time instant} $i$\\
$y(i)=\bar{\boldsymbol w}^H(i)\boldsymbol T_r^H(i)\boldsymbol
x(i)$;~~$e(i)=|y(i)|^2-1$\\
$\boldsymbol T_r(i+1)=\boldsymbol
T_r(i)-\mu_{T_r}e(i)y^{\ast}(i)\big[\boldsymbol I-\boldsymbol
a(\theta_0)\boldsymbol a^H(\theta_0)\big]\boldsymbol
x(i)\bar{\boldsymbol w}^H(i)$\\
$y(i)=\bar{\boldsymbol w}^H(i)\boldsymbol T_r^H(i+1)\boldsymbol
x(i)$;~~$e(i)=|y(i)|^2-1$\\
$\bar{\boldsymbol a}(\theta_0)=\boldsymbol T_r^H(i+1)\boldsymbol
a(\theta_0)$,~ $\bar{\boldsymbol x}(i)=\boldsymbol
T_r^H(i+1)\boldsymbol x(i)$\\
$\bar{\boldsymbol w}(i+1)=\bar{\boldsymbol
w}(i)-\mu_{\bar{w}}e(i)y^{\ast}(i)\big[\boldsymbol
I-\frac{\bar{\boldsymbol a}(\theta_0)\bar{\boldsymbol
a}^H(\theta_0)}{\bar{\boldsymbol a}^H(\theta_0)\bar{\boldsymbol
a}(\theta_0)}\big]\bar{\boldsymbol x}(i)$\\
\hline
    \end{tabular}
    \end{small}
\end{table}

\subsubsection{SG algorithm for the GSC}
For the GSC structure, assuming $\bar{\boldsymbol
w}_{\textrm{gsc}}(i)$ and $\boldsymbol T_{\textrm{gsc}}(i)$ are
given in (\ref{16}), respectively, we get
\begin{equation}\label{21}
\nabla
J_{\textrm{cm-gsc}_{T_{\textrm{gsc}}(i)}}(i)=e_{\textrm{gsc}}^{\ast}(i)\tilde{\boldsymbol
x}_{B}(i)\bar{\boldsymbol w}_{\textrm{gsc}}^H(i),
\end{equation}
\begin{equation}\label{22}
\nabla
J_{\textrm{cm-gsc}_{\bar{w}_{\textrm{gsc}}(i)}}(i)=e_{\textrm{gsc}}^{\ast}(i)\bar{\boldsymbol
x}_{B}(i),
\end{equation}
where $e_{\textrm{gsc}}(i)=1-\boldsymbol w^H(i)\tilde{\boldsymbol
x}(i)$ and $\boldsymbol w(i)$ is obtained from (\ref{9}).

Substituting (\ref{21}) and (\ref{22}) into the gradient rules, we
obtain the iterative SG algorithm for the GSC, which is summarized
in Table \ref{tab:JIO-CCM-SG-GSC}. where $\mu_{T_{\textrm{gsc}}}$
and $\mu_{\bar{w}_{\textrm{gsc}}}$ are the corresponding step size
factors for the GSC.
\begin{table}[!t]
\centering
    \caption{THE JIO-CCM-SG ALGORITHM FOR GSC}     % NOTE!  caption goes _before_ the table contents !!
    \label{tab:JIO-CCM-SG-GSC}
    \begin{small}
        \begin{tabular}{l}
\hline
\bfseries {Initialization:}\\
${\boldsymbol T}_{\textrm{gsc}}(1)=[{\boldsymbol I}_{r\times
r}~\boldsymbol 0_{r\times (m-r)}]^T$;
${\bar{\boldsymbol w}_{\textrm{gsc}}}(1)=\boldsymbol I_{r\times 1}$.\\
\bfseries {Update for each time instant} $i$\\
$\boldsymbol w(i)=\boldsymbol a_{\gamma}(\theta_0)-\boldsymbol
B^H\boldsymbol T_{\textrm{gsc}}(i)\bar{\boldsymbol
w}_{\textrm{gsc}}(i)$,~~
$y_{\textrm{gsc}}(i)=\boldsymbol w^H(i)\boldsymbol x(i)$\\
$\tilde{\boldsymbol x}(i)=y_{\textrm{gsc}}^{\ast}(i)\boldsymbol
x(i)$,~~ $\tilde{\boldsymbol x}_{B}(i)=\boldsymbol B
\tilde{\boldsymbol x}(i)$,~~
$e_{\textrm{gsc}}(i)=1-\boldsymbol w^H(i)\tilde{\boldsymbol x}(i)$\\
$\boldsymbol T_{\textrm{gsc}}(i+1)=\boldsymbol
T_{\textrm{gsc}}(i)-\mu_{T_r}e_{\textrm{gsc}}^{\ast}(i)\tilde{\boldsymbol
x}_{B}(i)\bar{\boldsymbol w}_{\textrm{gsc}}^H(i)$\\
$\boldsymbol w(i)=\boldsymbol a_{\gamma}(\theta_0)-\boldsymbol B^H\boldsymbol T_{\textrm{gsc}}(i+1)\bar{\boldsymbol w}_{\textrm{gsc}}(i)$\\
$y_{\textrm{gsc}}(i)=\boldsymbol w^H(i)\boldsymbol x(i)$,~~
$\tilde{\boldsymbol x}(i)=y_{\textrm{gsc}}^{\ast}(i)\boldsymbol
x(i)$,~~ $\tilde{\boldsymbol x}_{B}(i)=\boldsymbol B
\tilde{\boldsymbol x}(i)$\\
$e_{\textrm{gsc}}(i)=1-\boldsymbol w^H(i)\tilde{\boldsymbol
x}(i)$,~~ $\bar{\boldsymbol x}_{B}(i)=\boldsymbol
T_{\textrm{gsc}}^H(i+1)\tilde{\boldsymbol x}_{B}(i)$\\
$\bar{\boldsymbol w}_{\textrm{gsc}}(i+1)=\bar{\boldsymbol
w}_{\textrm{gsc}}(i)-\mu_{\bar{w}_{\textrm{gsc}}}e_{\textrm{gsc}}^{\ast}(i)\bar{\boldsymbol
x}_{B}(i)$\\
\hline
    \end{tabular}
    \end{small}
\end{table}

\subsection{Recursive Least Squares Algorithms}
In this part, we derive the RLS algorithms with the proposed CCM
reduced-rank scheme for both the DFP and the GSC structures.

\subsubsection{RLS algorithm for the DFP}
Considering the DFP case, the unconstrained least squares (LS) cost
function is given by
\begin{equation}\label{23}
\begin{split}
L_{\textrm{un}}\big(\boldsymbol T_r(i), \bar{\boldsymbol
w}(i)\big)&=\sum_{l=1}^{i}\alpha^{i-l}\big[|\bar{\boldsymbol
w}^H(i)\boldsymbol T_r^H(i)\boldsymbol x(l)|^2-1\big]^2\\
&+2\mathfrak{R}\Big\{\lambda\big[\bar{\boldsymbol w}^H(i)\boldsymbol
T_r^H(i)\boldsymbol a(\theta_0)-\gamma\big]\Big\},
\end{split}
\end{equation}
where $\alpha$ is a forgetting factor chosen as a positive constant
close to, but less than $1$.

Assuming $\bar{\boldsymbol w}(i)$ and $\boldsymbol T_r(i)$ are known
in (\ref{23}), respectively, we obtain
\begin{equation}\label{24}
\begin{split}
&\boldsymbol T_r(i+1)=\\
&\hat{\boldsymbol R}^{-1}(i)\Big\{\hat{\boldsymbol
p}(i)-\frac{\big[\hat{\boldsymbol p}^H(i)\hat{\boldsymbol
R}^{-1}(i)\boldsymbol a(\theta_0)-\gamma\big]\boldsymbol
a(\theta_0)}{\boldsymbol a^H(\theta_0)\hat{\boldsymbol
R}^{-1}(i)\boldsymbol a(\theta_0)}\Big\}\frac{\bar{\boldsymbol
w}^H(i)}{\|\bar{\boldsymbol w}(i)\|^2},
\end{split}
\end{equation}
\begin{equation}\label{25} \bar{\boldsymbol
w}(i+1)=\hat{\bar{\boldsymbol R}}^{-1}(i)\Big\{\hat{\bar{\boldsymbol
p}}(i)-\frac{\big[\hat{\bar{\boldsymbol
p}}^H(i)\hat{\bar{\boldsymbol R}}^{-1}(i)\bar{\boldsymbol
a}(\theta_0)-\gamma\big]\bar{\boldsymbol
a}(\theta_0)}{\bar{\boldsymbol a}^H(i)\hat{\bar{\boldsymbol
R}}^{-1}(i)\bar{\boldsymbol a}(\theta_0)}\Big\},
\end{equation}
where $\hat{\boldsymbol
R}(i)=\sum_{l=1}^{i}\alpha^{i-l}|y(l)|^2\boldsymbol x(l)\boldsymbol
x^H(l)$, $\hat{\bar{\boldsymbol
R}}(i)=\sum_{l=1}^{i}\alpha^{i-l}|y(l)|^2\bar{\boldsymbol
x}(l)\bar{\boldsymbol x}^H(l)$, $\hat{\boldsymbol
p}(i)=\sum_{l=1}^{i}\alpha^{i-l}y^{\ast}(l)\boldsymbol x(l)$, and
$\hat{\bar{\boldsymbol p}}(i)=\sum_{l=1}^{i}\alpha^{i-l}
y^{\ast}(l)\bar{\boldsymbol x}(l)$ with $y(i)$ expressed in
(\ref{10}). {  The derivation of (\ref{24}) is given in the
appendix}. Note that $\hat{\boldsymbol R}(i)$ is not invertible if
$i<m$. It can be implemented by employing the diagonal loading
technique \cite{Trees}, \cite{Jian}. This same procedure is also
used for the remaining matrices.

To avoid the matrix inversion and reduce the complexity, we employ
the matrix inversion lemma \cite{Haykin} to update $\hat{\boldsymbol
R}^{-1}(i)$ and $\hat{\bar{\boldsymbol R}}(i)$ iteratively. { The
resulting adaptive algorithm, which we denominate as JIO-CCM-RLS, is
summarized in Table \ref{tab:JIO-CCM-RLS-DFP}}, where
$\hat{\boldsymbol\Phi}(i)=\hat{\boldsymbol R}^{-1}(i)$ and
$\hat{\bar{\boldsymbol\Phi}}(i)=\hat{\bar{\boldsymbol R}}^{-1}(i)$
are defined for concise presentation, $\boldsymbol k(i)\in\mathbb
C^{m\times1}$ and $\bar{\boldsymbol k}(i)\in\mathbb C^{r\times1}$
are the full-rank and reduced-rank gain vectors, respectively. The
recursive procedures are implemented by initializing
$\hat{{\boldsymbol\Phi}}(0)={\delta}\boldsymbol I_{m\times m}$ and
$\hat{\bar{\boldsymbol\Phi}}(0)=\bar{\delta}\boldsymbol I_{r\times
r}$, where $\delta$ and $\bar{\delta}$ are positive scalars.

\begin{table}[!t]
\centering
    \caption{THE JIO-CCM-RLS ALGORITHM FOR DFP}     % NOTE!  caption goes _before_ the table contents !!
    \label{tab:JIO-CCM-RLS-DFP}
    \begin{small}
        \begin{tabular}{l}
\hline
\bfseries {Initialization:}\\
${\boldsymbol T}_r(1)=[{\boldsymbol I}_{r\times r}~\boldsymbol 0_{r\times (m-r)}]^T$;\\
${\bar{\boldsymbol w}}(1)=\boldsymbol T_r^H(1)\boldsymbol a_{\gamma}(\theta_0)/(\|\boldsymbol T_r^H(1)\boldsymbol a_{\gamma}(\theta_0)\|^2)$;\\
$\hat{\boldsymbol\Phi}(0)=\delta\boldsymbol I_{m\times
m}$,$\hat{\bar{\boldsymbol\Phi}}(0)=\bar{\delta}\boldsymbol
I_{r\times r}$,$\hat{\boldsymbol p}(0)=\boldsymbol 0_{m\times 1}$,$\hat{\bar{\boldsymbol p}}(0)=\boldsymbol 0_{r\times 1}$.\\
\bfseries {Update for each time instant} $i$\\
$y(i)=\bar{\boldsymbol w}^H(i)\boldsymbol T_r^H(i)\boldsymbol
x(i)$,~~$\hat{\boldsymbol p}(i)=\alpha\hat{\boldsymbol p}(i-1)+y^{\ast}(i)\boldsymbol x(i)$\\
$\boldsymbol
k(i)=\frac{\alpha^{-1}\hat{\boldsymbol\Phi}(i-1)\boldsymbol
x(i)}{(1/|y(i)|^2)+\alpha^{-1}\boldsymbol x^H(i)\hat{\boldsymbol
\Phi}(i-1)\boldsymbol x(i)}$\\
$\hat{\boldsymbol\Phi}(i)=\alpha^{-1}\hat{\boldsymbol
\Phi}(i-1)-\alpha^{-1}\boldsymbol k(i)\boldsymbol
x^H(i)\hat{\boldsymbol\Phi}(i-1)$\\
$\boldsymbol T_r(i+1)=\hat{\boldsymbol
\Phi}(i)\Big\{\hat{\boldsymbol p}(i)-\frac{\big[\hat{\boldsymbol
p}^H(i)\hat{\boldsymbol\Phi}(i)\boldsymbol
a(\theta_0)-\gamma\big]\boldsymbol a(\theta_0)}{\boldsymbol
a^H(\theta_0)\hat{\boldsymbol\Phi}(i)\boldsymbol
a(\theta_0)}\Big\}\frac{\bar{\boldsymbol
w}^H(i)}{\|\bar{\boldsymbol w}(i)\|^2}$\\
$y(i)=\bar{\boldsymbol w}^H(i)\boldsymbol T_r^H(i+1)\boldsymbol
x(i)$,~$\bar{\boldsymbol a}(\theta_0)=\boldsymbol
T_r^H(i+1)\boldsymbol
a(\theta_0)$\\
$\bar{\boldsymbol x}(i)=\boldsymbol
T_r^H(i+1)\boldsymbol x(i)$,~$\hat{\bar{\boldsymbol p}}(i)=\alpha\hat{\bar{\boldsymbol p}}(i-1)+y^{\ast}(i)\bar{\boldsymbol x}(i)$\\
$\bar{\boldsymbol
k}(i)=\frac{\alpha^{-1}\hat{\bar{\boldsymbol\Phi}}(i-1)\bar{\boldsymbol
x}(i)}{(1/|y(i)|^2)+\alpha^{-1}\bar{\boldsymbol
x}^H(i)\hat{\bar{\boldsymbol \Phi}}(i-1)\bar{\boldsymbol x}(i)}$\\
$\hat{\bar{\boldsymbol\Phi}}(i)=\alpha^{-1}\hat{\bar{\boldsymbol
\Phi}}(i-1)-\alpha^{-1}\bar{\boldsymbol k}(i)\bar{\boldsymbol
x}^H(i)\hat{\bar{\boldsymbol\Phi}}(i-1)$\\
$\bar{\boldsymbol w}(i+1)=\hat{\bar{\boldsymbol
\Phi}}(i)\big\{\hat{\bar{\boldsymbol
p}}(i)-\frac{\big[\hat{\bar{\boldsymbol
p}}^H(i)\hat{\bar{\boldsymbol\Phi}}(i)\bar{\boldsymbol
a}(\theta_0)-\gamma\big]\bar{\boldsymbol
a}(\theta_0)}{\bar{\boldsymbol a}^H(i)\hat{\bar{\boldsymbol\Phi}}(i)\bar{\boldsymbol a}(\theta_0)}\big\}$\\
\hline
    \end{tabular}
    \end{small}
\end{table}

\subsubsection{RLS algorithm for the GSC}
For the GSC structure, the LS cost function is given by
\begin{equation}\label{26}
\begin{split}
&L_{\textrm{un-gsc}}(\boldsymbol T_{\textrm{gsc}}(i),
\bar{\boldsymbol
w}_{\textrm{gsc}}(i))=\\
&~~~\sum_{l=1}^{i}\alpha^{i-l}\Big\{\big[\boldsymbol
a_{\gamma}(\theta_0)-\boldsymbol B^H\boldsymbol
T_{\textrm{gsc}}(i)\bar{\boldsymbol
w}_{\textrm{gsc}}(i)\big]^H\tilde{\boldsymbol x}(l)-1\Big\}^2.
\end{split}
\end{equation}

Assuming the optimal reduced-rank weight vector $\bar{\boldsymbol
w}_{\textrm{gsc}}$ and the transformation matrix $\boldsymbol
T_{\textrm{gsc}}(i)$ are known, respectively, computing the
gradients of (\ref{26}) with respect to $\boldsymbol
T_{\textrm{gsc}}(i)$ and $\bar{\boldsymbol w}_{\textrm{gsc}}(i)$,
and equating them equal to null, we have
\begin{equation}\label{27}
\boldsymbol T_{\textrm{gsc}}(i+1)=\hat{\boldsymbol
R}_{\tilde{x}_{B}}^{-1}(i)\hat{\tilde{\boldsymbol
p}}_{B}(i)\frac{\bar{\boldsymbol
w}_{\textrm{gsc}}^H(i)}{\|\bar{\boldsymbol
w}_{\textrm{gsc}}(i)\|^2},
\end{equation}
\begin{equation}\label{28}
\bar{\boldsymbol w}_{\textrm{gsc}}(i+1)=\hat{\bar{\boldsymbol
R}}_{\bar{x}_{{B}}}^{-1}(i)\hat{\bar{\boldsymbol p}}_{{B}}(i),
\end{equation}
where $\hat{\boldsymbol R}_{\tilde{
x}_{B}}(i)=\sum_{l=1}^{i}\alpha^{i-l}\boldsymbol B\tilde{\boldsymbol
x}(l)\tilde{\boldsymbol x}^H(l)\boldsymbol B^H$,
$\hat{\bar{\boldsymbol R}}_{\bar{x}_{B}}(i)=\boldsymbol
T_{\textrm{gsc}}^H(i)\hat{\boldsymbol
R}_{\tilde{x}_{{B}}}(i)\boldsymbol T_{\textrm{gsc}}(i)$,
$\hat{\tilde{\boldsymbol
p}}_{B}(i)=\sum_{l=1}^{i}\alpha^{i-l}\boldsymbol
[d_0^{\ast}(l)-1]\tilde{\boldsymbol x}_{B}(l)$, and
$\hat{\bar{\boldsymbol p}}_{{B}}(i)=\boldsymbol
T_{\textrm{gsc}}^H(i)\hat{\tilde{\boldsymbol p}}_{{B}}(i)$.

Setting $\hat{\boldsymbol\Phi}_{\tilde{ x}_{B}}(i)=\hat{\boldsymbol
R}_{\tilde{ x}_{B}}^{-1}(i)$ and
$\hat{\bar{\boldsymbol\Phi}}_{\bar{x}_{{B}}}(i)=\hat{\bar{\boldsymbol
R}}_{\bar{x}_{{B}}}^{-1}(i)$ and employing the matrix inversion
lemma yields,
\begin{equation}\label{29}
\boldsymbol T_{\textrm{gsc}}(i+1)=\boldsymbol
T_{\textrm{gsc}}(i)-{\boldsymbol k}_{{B}}(i)\boldsymbol
e_{\bar{w}_{\textrm{gsc}}}(i),
\end{equation}
\begin{equation}\label{30}
\bar{\boldsymbol w}_{\textrm{gsc}}(i+1)=\bar{\boldsymbol
w}_{\textrm{gsc}}(i)-e_{\textrm{gsc}}^{\ast}(i)\bar{\boldsymbol
k}_{{B}}(i),
\end{equation}
where $\boldsymbol k_B(i)\in\mathbb C^{(m-1)\times1}$ and
$\bar{\boldsymbol k}_B\in\mathbb C^{r\times1}$ are gain vectors,
$\boldsymbol e_{\bar{w}_{\textrm{gsc}}}(i)=[1-\tilde{\boldsymbol
x}^H(i)\boldsymbol w(i)]\frac{\bar{\boldsymbol
w}_{\textrm{gsc}}^H(i)}{\|\bar{\boldsymbol
w}_{\textrm{gsc}}(i)\|^2}$, $e_{\textrm{gsc}}(i)=1-\boldsymbol
w^H(i)\tilde{\boldsymbol x}(i)$, and $\boldsymbol w(i)$ is defined
by (\ref{9}). A summary of the reduced-rank RLS algorithm with the
CCM design for the GSC is given in Table \ref{tab:JIO-CCM-RLS-GSC}.

\begin{table}[!t]
\centering
    \caption{THE JIO-CCM-RLS ALGORITHM FOR GSC}     % NOTE!  caption goes _before_ the table contents !!
    \label{tab:JIO-CCM-RLS-GSC}
    \begin{small}
        \begin{tabular}{l}
\hline
\bfseries {Initialization:}\\
${\boldsymbol T}_{\textrm{gsc}}(1)=[{\boldsymbol I}_{r\times
r}~\boldsymbol 0_{r\times (m-r)}]^T$,~~${\bar{\boldsymbol
w}}_{\textrm{gsc}}(1)=\boldsymbol I_{r\times 1}$\\
$\hat{\boldsymbol\Phi}_{{\tilde{x}_{{B}}}}(0)=\delta\boldsymbol
I_{m\times
m}$,~~$\hat{\bar{\boldsymbol\Phi}}_{\bar{x}_{{B}}}(0)=\bar{\delta}\boldsymbol
I_{r\times r}$.\\
\bfseries {Update for each time instant} $i$\\
$\boldsymbol w(i)=\boldsymbol a_{\gamma}(\theta_0)-\boldsymbol
B^H\boldsymbol T_{\textrm{gsc}}(i)\bar{\boldsymbol
w}_{\textrm{gsc}}(i)$,~~
$y_{\textrm{gsc}}(i)=\boldsymbol w^H(i)\boldsymbol x(i)$\\
$\tilde{\boldsymbol x}(i)=y_{\textrm{gsc}}^{\ast}(i)\boldsymbol
x(i)$,~~$\tilde{\boldsymbol x}_{{B}}(i)=\boldsymbol
B\tilde{\boldsymbol x}(i)$,~~$e_{\textrm{gsc}}(i)=1-\boldsymbol
w^H(i)\tilde{\boldsymbol x}(i)$\\
${\boldsymbol
k}_{{B}}(i)=\frac{\alpha^{-1}\hat{\boldsymbol\Phi}_{\tilde{
x}_{{B}}}(i-1)\tilde{\boldsymbol
x}_{{B}}(i)}{1+\alpha^{-1}\tilde{\boldsymbol
x}_{{B}}^H(i)\hat{\boldsymbol\Phi}_{\tilde{
x}_{{B}}}(i-1)\tilde{\boldsymbol x}_{{B}}(i)}$\\
$\hat{\boldsymbol\Phi}_{\tilde{x}_{{B}}}(i)=\alpha^{-1}\hat{\boldsymbol\Phi}_{\tilde{x}_{{B}}}(i-1)-\alpha^{-1}{\boldsymbol
k}_{{B}}(i)\tilde{\boldsymbol
x}_{{B}}^H(i)\hat{\boldsymbol\Phi}_{\tilde{
x}_{{B}}}(i-1)$\\
$\boldsymbol e_{\bar{w}_{\textrm{gsc}}}(i)=[1-\tilde{\boldsymbol
x}^H(i)\boldsymbol w(i)]\frac{\bar{\boldsymbol
w}_{\textrm{gsc}}^H(i)}{\|\bar{\boldsymbol
w}_{\textrm{gsc}}(i)\|^2}$\\
$\boldsymbol T_{\textrm{gsc}}(i+1)=\boldsymbol
T_{\textrm{gsc}}(i)-{\boldsymbol
k}_{{B}}(i)\boldsymbol e_{\bar{w}_{\textrm{gsc}}}(i)$\\
$\bar{\boldsymbol x}_{{B}}(i)=\boldsymbol
T_{\textrm{gsc}}^H(i+1)\tilde{\boldsymbol x}_{{B}}(i)$\\
$\bar{\boldsymbol
k}_{{B}}(i)=\frac{\alpha^{-1}\hat{\bar{\boldsymbol\Phi}}_{\bar{
x}_{{B}}}(i-1)\bar{\boldsymbol
x}_{{B}}(i)}{1+\alpha^{-1}\bar{\boldsymbol
x}_{{B}}^H(i)\hat{\bar{\boldsymbol\Phi}}_{\bar{
x}_{{B}}}(i-1)\bar{\boldsymbol x}_{{B}}(i)}$\\
$\hat{\bar{\boldsymbol\Phi}}_{\bar{x}_{{B}}}(i)=\alpha^{-1}\hat{\bar{\boldsymbol\Phi}}_{\bar{x}_{{B}}}(i-1)-\alpha^{-1}\bar{\boldsymbol
k}_{{B}}(i)\bar{\boldsymbol
x}_{{B}}^H(i)\hat{\bar{\boldsymbol\Phi}}_{\bar{
x}_{{B}}}(i-1)$\\
$\boldsymbol w(i)=\boldsymbol a_{\gamma}(\theta_0)-\boldsymbol
B^H\boldsymbol
T_{\textrm{gsc}}(i+1)\bar{\boldsymbol w}_{\textrm{gsc}}(i)$\\
$e_{\textrm{gsc}}(i)=1-\boldsymbol w^H(i)\tilde{\boldsymbol x}(i)$\\
$\bar{\boldsymbol w}_{\textrm{gsc}}(i+1)=\bar{\boldsymbol
w}_{\textrm{gsc}}(i)-e_{\textrm{gsc}}^{\ast}(i)\bar{\boldsymbol
k}_{{B}}(i)$\\
\hline
    \end{tabular}
    \end{small}
\end{table}

\subsection{Gram-Schmidt Technique for Problem 2}

As mentioned before, the transformation matrix $\boldsymbol
T_r(i+1)$ for the DFP is constituted by a bank of full-rank filters
$\boldsymbol t_{j}(i+1)~(j=1,\ldots, r)$, {  which cannot be
guaranteed to be orthogonal. According to the optimization problem 2
in (\ref{12}), the transformation matrix $\boldsymbol T_r(i)$ can be
reformulated to compose $r$ orthogonal vectors, which span the same
subspace generated by the original vectors. The reformulation
ensures that the projection of the received vector onto each
dimension is one time and avoids the overlap (e.g., takes the same
information twice or more). Compared with the original
transformation matrix, the reformulated transformation matrix is
more efficient to keep the useful information in the generated
reduced-rank received vector for the parameter estimation}. The
orthogonal procedure is realized by the Gram-Schmidt (GS) technique
\cite{Golub}. Specifically, after the iterative processes for the
computation of the transformation matrix, the GS technique is
performed to modify the columns of the transformation matrix as
follows:
\begin{equation}\label{31}
\boldsymbol t_{j,\textrm{ort}}(i+1)=\boldsymbol
t_{j}(i+1)-\sum_{l=1}^{j-1}\textrm{proj}_{{\boldsymbol
t}_{l,\textrm{ort}}(i+1)}\boldsymbol t_j(i+1),
\end{equation}
where $\boldsymbol t_{j,\textrm{ort}}(i+1)$ is the normalized
orthogonal vector after the GS process. The projection operator is
$\textrm{proj}_{{\boldsymbol t}_{l,\textrm{ort}}(i+1)}\boldsymbol
t_j(i+1)=[\boldsymbol t_{l,\textrm{ort}}^H(i+1)\boldsymbol
t_{l,\textrm{ort}}(i+1)]^{-1}[\boldsymbol
t_{l,\textrm{ort}}^H(i+1)\boldsymbol t_j(i+1)]\boldsymbol
t_{l,\textrm{ort}}(i+1)$.

The reformulated transformation matrix $\boldsymbol
T_{r,\textrm{ort}}(i+1)$ is constructed after we obtain a set of
orthogonal $\boldsymbol t_{j,\textrm{ort}}(i+1)$. By employing
$\boldsymbol T_{r,\textrm{ort}}(i+1)$ to compute the reduced-rank
weight vectors, the adaptive algorithms
%get the following quantities in Table \ref{tab:JIO-CCM-SG-DFP} and
%Table \ref{tab:JIO-CCM-SG-GSC}, and update the reduced-rank weight
%vector $\bar{\boldsymbol w}(i)$ in (\ref{26}) and $\bar{\boldsymbol
%w}_{\textrm{gsc}}(i)$ in (\ref{30}), respectively, the proposed
could achieve an improved performance. Following the same
procedures, we can also apply the GS technique to the adaptive
algorithms for the GSC structure. Simulations will be given to show
this result. We denominate the GS version of the SG and RLS
algorithms as JIO-CCM-GS and JIO-CCM-RGS, respectively.

\subsection{Automatic Rank Selection}
The selection of the rank $r$ impacts the performance of the
proposed reduced-rank algorithms. Here, we introduce an adaptive
method for selecting the rank. Related works on the rank selection
for the MSWF and the AVF techniques have been reported in
\cite{Honig} and \cite{Qian}, respectively. Unlike these methods, we
describe a rank selection method based on the CM criterion computed
by the filters $\boldsymbol T_r^{(r)}(i)$ and $\bar{\boldsymbol
w}^{(r)}(i)$, where the superscript $(\cdot)^{(r)}$ denotes the rank
used for the adaptation at each time instant. We consider the rank
adaptation technique for both the DFP and the GSC structures.
Specifically, in the DFP structure, the rank is automatically
selected for the proposed algorithms based on the
exponentially-weighted a \textit{posteriori} least-squares cost
function according to the CM criterion, which is
\begin{equation}\label{32}
\begin{split}
&J_{\textrm{pcm}}\big(\boldsymbol T_r^{(r)}(i-1), \bar{\boldsymbol
w}^{(r)}(i-1)\big)=\\
&~~~~~~~~~~\sum_{l=1}^{i}\varrho^{i-l}\big[|\bar{\boldsymbol
w}^{(r)H}(l-1)\boldsymbol T_r^{(r)}(l-1)\boldsymbol
x(l)|^2-1\big]^2,
\end{split}
\end{equation}
where $\varrho$ is the exponential weight factor that is required as
the optimal rank $r$ can change as a function of the time instant
$i$. From the expressions in Table \ref{tab:JIO-CCM-SG-DFP} and
Table \ref{tab:JIO-CCM-RLS-DFP}, the key quantities to be updated
for the rank adaptation are the transformation matrix $\boldsymbol
T_r(i)$, the reduced-rank weight vector $\bar{\boldsymbol w}(i)$,
the associated reduced-rank steering vector $\bar{\boldsymbol
a}(\theta_0)$ and the matrix $\hat{\bar{\boldsymbol\Phi}}(i)$ (for
RLS only). To this end, we express $\boldsymbol T_r^{(r)}(i)$ and
$\bar{\boldsymbol w}^{(r)}(i)$ for the rank adaptation as follows:
\begin{equation}\label{33}
\begin{split}
&\boldsymbol T_r^{(r)}(i)=\begin{bmatrix}
 t_{1,1} & t_{1,2} & \ldots & t_{1,r_{\textrm{min}}} & \ldots &
 t_{1,r_{\textrm{max}}}\\
 t_{2,1} & t_{2,2} & \ldots & t_{2,r_{\textrm{min}}} & \ldots &
 t_{2,r_{\textrm{max}}}\\
 \vdots & \vdots & \vdots & \vdots & \vdots & \vdots\\
 t_{m,1} & t_{m,2} & \ldots & t_{m,r_{\textrm{min}}} & \ldots &
 t_{m,r_{\textrm{max}}}\\ \end{bmatrix},\\
&\bar{\boldsymbol w}^{(r)}(i)=\begin{bmatrix}
 \bar{w}_{1} & \bar{w}_{2} & \ldots & \bar{w}_{r_{\textrm{min}}} & \ldots &
 \bar{w}_{r_{\textrm{max}}}\\ \end{bmatrix}^{T},
\end{split}
\end{equation}
where $r_{\textrm{min}}$ and $r_{\textrm{max}}$ are the minimum and
maximum ranks allowed, respectively.

For each time instant $i$, $\boldsymbol T_r^{(r)}(i)$ and
$\bar{\boldsymbol w}^{(r)}(i)$ are updated along with the associated
quantities $\bar{\boldsymbol a}(\theta_0)$ and
$\hat{\bar{\boldsymbol\Phi}}(i)$ for a selected $r$ according to the
minimization of the cost function in (\ref{32}). The developed
automatic rank selection method is given by
\begin{equation}\label{34}
r_{\textrm{opt}}=\arg\min_{r_{\textrm{min}}\leq j\leq
r_{\textrm{max}}}J_{\textrm{pcm}}\big(\boldsymbol T_r^{(j)}(i-1),
\bar{\boldsymbol w}^{(j)}(i-1)\big),
\end{equation}
where $j$ is an integer ranging between $r_{\textrm{min}}$ and
$r_{\textrm{max}}$. Note that a smaller rank may provide faster
adaptation during the initial stages of the estimation procedure and
a slightly larger rank tends to yield a better steady-state
performance. Our studies reveal that the range for which the rank
$r$ of the proposed algorithms have a positive impact on the
performance is very limited, being from $r_{\textrm{min}}=3$ to
$r_{\textrm{max}}=7$. These values are rather insensitive to the
number of users in the system, to the number of sensor elements, and
work efficiently for the studied scenarios. The additional
complexity of this automatic rank selection technique is for the
update of involved quantities with the maximum allowed rank
$r_{\textrm{max}}$ and the computation of the cost function in
(\ref{32}). With the case of large $m$, the rank $r$ is
significantly smaller than $m$ and the additional computations do
not increase the computational cost significantly.

The proposed algorithms with the rank adaptation technique can
increase the convergence rate and improve the output performance,
and $r$ can be made fixed once the algorithms reach the
steady-state. Simulation results will show how the developed rank
adaptation technique works. Note that the same idea can be employed
in the algorithms for the GSC structure. We omit this part for
simplification and readability.

\section{Analysis of the proposed algorithms}
In this section, we provide a complexity analysis of the proposed
reduced-rank algorithms and compare them with existing algorithms.
An analysis of the optimization problem for the proposed
reduced-rank scheme is also carried out.

\subsection{Complexity Analysis}
We evaluate the computational complexity of the proposed
reduced-rank algorithms and compare them with the existing full-rank
and reduced-rank algorithms based on the MSWF and the AVF techniques
for the DFP and the GSC structures. With respect to each algorithm,
we consider the CMV and the CCM design criteria. The computational
requirements are described in terms of the number of complex
arithmetic operations, namely, additions and multiplications. The
complexity of the proposed and existing algorithms for the DFP is
depicted in Table \ref{tab: Complexity_DFP} and for the GSC in Table
\ref{tab: Complexity_GSC}. Since we did not consider the AVF
technique for the GSC structure, we put its complexity for the DFP
in both tables for comparison.

For the DFP structure, we can say that the complexity of the
proposed reduced-rank SG type and extended GS version algorithms
increases linearly with $rm$. The parameter $m$ is more influential
since $r$ is selected around a small range that is much less than
$m$ for large arrays. The complexity of the proposed reduced-rank
RLS type and GS version algorithms is higher than that of the SG
type and quadratic with $m$ and $r$. For the GSC structure, the
complexity of the SG type algorithms has extra $m^2$ terms as
compared to the DFP structure in terms of additions and
multiplications due to the blocking matrix in the sidelobe
canceller. There is no significant difference in complexity of the
RLS type algorithms due to the presence of the blocking matrix since
(\ref{29}) and (\ref{30}) are recursive expressions and, as compared
to non-recursive versions, reduce the complexity.

\begin{figure}[!htb]
\begin{center}
\def\epsfsize#1#2{1.0\columnwidth}
\epsfbox{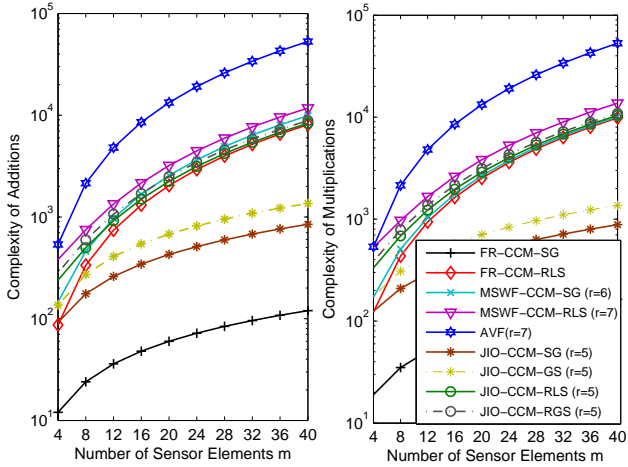} \caption{Complexity in terms of arithmetic
operations versus the length of the filter $m$ for the DFP
structure.} \label{fig:complexity_add_mul_dsp_gram_final3}
\end{center}
\end{figure}

\begin{figure}[!htb]
\begin{center}
\def\epsfsize#1#2{1.0\columnwidth}
\epsfbox{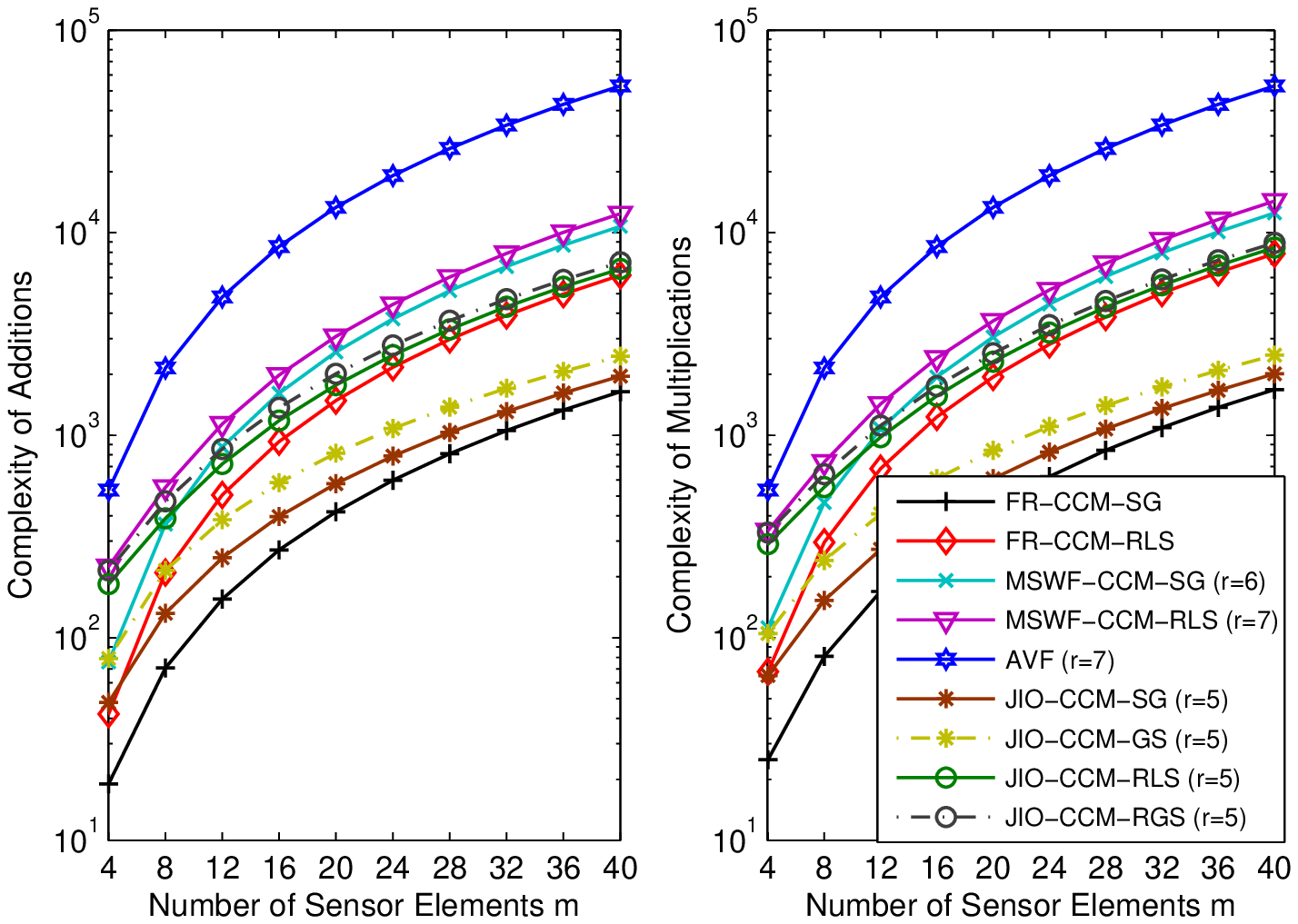} \caption{Complexity in terms of arithmetic
operations versus the length of the filter $m$ for the GSC
structure.} \label{fig:complexity_add_mul_gsc_gram_final3}
\end{center}
\end{figure}

In order to illustrate the main trends in what concerns the
complexity of the proposed algorithms, we show in Fig.
\ref{fig:complexity_add_mul_dsp_gram_final3} and Fig.
\ref{fig:complexity_add_mul_gsc_gram_final3} the complexity of both
the DFP and the GSC structures in terms of additions and
multiplications versus the length of the filter $m$. Since the
complexity of the current algorithms according to the CMV criterion
is a little less than that of the CCM criterion, we only plot the
curves for the CCM criterion for simplification. {  Note that the
values of $r$ are different with respect to different algorithms,
which are set to make the corresponding algorithms reach the best
performance according to the experiments. The specific values are
given in the figures}. It is clear that the proposed SG type and
extended GS version algorithms have a complexity slightly higher
than the full-rank SG algorithm but much less than the existing
algorithms based on the MSWF and the AVF techniques for both the DFP
and the GSC structures. The curves of the proposed RLS type and GS
version algorithms are situated between the full-rank RLS and the
MSWF RLS algorithms in both figures.

\begin{table}
\centering \caption{\normalsize Computational complexity of
algorithms for DFP} \footnotesize \label{tab: Complexity_DFP}
\begin{tabular}{l c c}
\hline \\
Algorithm & Additions                   & Multiplications\\
\hline   \\
FR-CMV-SG    & $3m-1$                        & $4m+1$ \\
FR-CCM-SG    & $3m$                  & $4m+3$ \\
FR-CMV-RLS   & $4m^2-m-1$            & $5m^2+5m-1$\\
FR-CCM-RLS   & $5m^2+2m-1$              & $6m^2+6m+3$\\

MSWF-CMV-SG         & $rm^2+(r+1)m$             & $rm^{2}+2rm$ \\
                 & $+2r-2$                            & $+5r+2$\\
MSWF-CCM-SG        & $rm^2+(r+1)m$             & $rm^2+2rm$ \\
                & $+4r-2$                 & $+4r+4$\\
MSWF-CMV-RLS    & $rm^2+(r+1)m$   & $(r+1)m^2+2rm$\\
                & $+4r^2-3r-1$  & $+5r^2+4r$\\
MSWF-CCM-RLS    & $rm^2+(r+1)m$ & $(r+1)m^2+2rm$\\
                & $+5r^2-r$        & $+6r^2+7r+3$\\
AVF           & $(4r+5)m^2+(r-1)m$   & $(5r+8)m^2$\\
              & $-2r-1$     & $+(3r+2)m$\\
JIO-CMV-SG        & $4rm+m+2r-3$            & $4rm+m+7r+3$\\
JIO-CMV-GS     & $7rm-m-1$               & $7rm-2m+8r+2$\\
JIO-CCM-SG        & $4rm+m+2r-2$             & $4rm+m+7r+6$\\
JIO-CCM-GS     & $7rm-m$                  & $7rm-2m+8r+5$\\
JIO-CMV-RLS    & $4m^2+(2r-1)m$           & $5m^2+(3r+3)m$\\
               & $+4r^2-4r-1$              & $+6r^2+4r$\\
JIO-CMV-RGS    & $4m^2+(5r-3)m$            & $5m^2+6rm$\\
               & $+4r^2-6r+1$              & $6r^2+5r-1$\\
JIO-CCM-RLS    & $5m^2+rm$                & $6m^2+(2r+6)m$\\
               & $+5r^2+3r-1$              & $+5r^2+9r+3$\\
JIO-CCM-RGS    & $5m^2+(4r-2)m$            & $6m^2+(5r+3)m$\\
               & $5r^2+r+1$                & $5r^2+10r+2$\\
\hline
\end{tabular}
\end{table}

\begin{table}
\centering \caption{\normalsize Computational complexity of
algorithms for GSC} \footnotesize \label{tab: Complexity_GSC}
\begin{tabular}{l c c}
\hline \\
Algorithm & Additions                   & Multiplications\\
\hline   \\
FR-CMV-SG    & $m^2+m-2$                        & $m^2+2m-1$ \\
FR-CCM-SG    & $m^2+m-1$                  & $m^2+2m+1$ \\
FR-CMV-RLS   & $4m^2-6m+4$            & $5m^2-4m$\\
FR-CCM-RLS   & $4m^2-6m+2$              & $5m^2-3m$\\

MSWF-CMV-SG         & $(r+1)m^2-2rm$             & $(r+2)m^{2}-(r+2)m$ \\
                 & $+2r-1$                            & $+2r+2$\\
MSWF-CCM-SG        & $(r+1)m^2-2rm+2r$             & $(r+2)m^2-(r+2)m$ \\
                &                  & $+2r+4$\\
MSWF-CMV-RLS    & $(r+1)m^2-2rm$   & $(r+2)m^2-(r+2)m$\\
                & $+3r^2+r-1$  & $+4r^2+4r$\\
MSWF-CCM-RLS    & $(r+1)m^2-2rm$ & $(r+2)m^2-(r+1)m$\\
                & $+3r^2+r-1$        & $+4r^2+4r+1$\\
AVF           & $(4r+5)m^2+(r-1)m$   & $(5r+8)m^2$\\
              & $-2r-1$     & $+(3r+2)m$\\
JIO-CMV-SG        & $m^2+2rm-m-r$            & $m^2+2rm+r+2$\\
JIO-CMV-GS     & $m^2+5rm-3m$               & $m^2+5rm-3m$\\
               & $-6r+4$                    & $-r+4$\\
JIO-CCM-SG        & $m^2+2rm-m-r+1$             & $m^2+2rm+r+4$\\
JIO-CCM-GS     & $m^2+5rm-3m$                  & $m^2+5rm-3m$\\
               & $-6r+5$                       & $-r+6$\\
JIO-CMV-RLS    & $4m^2+(2r-8)m$           & $5m^2+(2r-6)m$\\
               & $+5r^2-2r+4$              & $+7r^2+3r+2$\\
JIO-CMV-RGS    & $4m^2+(5r-10)m$            & $5m^2+(5r-9)m$\\
               & $+5r^2-7r+8$              & $7r^2+r+4$\\
JIO-CCM-RLS    & $4m^2+(2r-7)m$                & $5m^2+(2r-4)m$\\
               & $+5r^2-4r+3$              & $+7r^2+2r+1$\\
JIO-CCM-RGS    & $4m^2+(5r-9)m$            & $5m^2+(5r-7)m$\\
               & $5r^2-9r+7$                & $7r^2+3$\\
\hline
\end{tabular}
\end{table}

\subsection{Analysis of the Optimization Problem}
Here, we present the analysis of the proposed reduced-rank scheme
according to the CCM criterion, which depends on the transformation
matrix and the reduced-rank weight vector. {  Our approach starts
from the analysis of the full-rank constant modulus criterion and
then utilizes the transformation matrix and the reduced-rank weight
vector with the received vector to express the output. The
constraint is enforced during the analysis. We will consider the
analysis for both the DFP and the GSC structures}.

{  The full-rank constant modulus cost function in (\ref{2}) with
$p=2$ and $\nu=1$ can be written as
\begin{equation}\label{35}
\begin{split}
J_{\textrm{cm}}\big(\boldsymbol w(i)\big)&=\mathbb
E\big[|y(i)|^4-2|y(i)|^2+1\big]\\
&=\mathbb E\big[|\boldsymbol w^H(i)\boldsymbol x(i)\boldsymbol
x^H(i)\boldsymbol w(i)|^2\big]\\
&~~~-2\mathbb E\big[|\boldsymbol w^H(i)\boldsymbol x(i)|^2\big]+1,
\end{split}
\end{equation}
where $\boldsymbol x(i)=\sum_{k=0}^{q-1}C_k d_k(i)\boldsymbol
a(\theta_k)+\boldsymbol n(i)$ ($k=0, \ldots, q-1$) from (\ref{1})
with $D_k$ being the signal amplitude and $d_k$ is the transmitted
bit of the $k$th user. Note that we have replaced $s_k$ in (\ref{1})
by $C_k d_k$.

For the sake of analysis, we will follow the assumption in \cite{Xu}
and consider a noise free case. For small noise variance
$\sigma_n^2$, this assumption can be considered as a small
perturbation and the analysis will still be applicable. For large
$\sigma_n^2$, we remark that the term $\gamma$ can be adjusted for
the analysis. Under this assumption, we write the received vector as
$\boldsymbol x(i)=\boldsymbol A(\boldsymbol\theta)\boldsymbol
C\boldsymbol d(i)$, where $\boldsymbol A(\boldsymbol\theta)$, as
before, denotes the signature matrix, $\boldsymbol
C(i)=\textrm{diag}[C_0, \ldots, C_{q-1}]\in\mathbb C^{q\times q}$,
and $\boldsymbol d(i)=[d_0(i), \ldots, d_{q-1}(i)]^T\in\mathbb
C^{q\times1}$.

For simplicity, we drop the time instant in the quantities. Letting
$\varsigma_k=C_k\boldsymbol w^H\boldsymbol a(\theta_k)$ and
$\boldsymbol\varsigma=[\varsigma_0, \ldots, \varsigma_{q-1}]^T$, we
have
\begin{equation}\label{36}
J_{\textrm{cm}}=\mathbb E[\boldsymbol\varsigma^H\boldsymbol
d\boldsymbol
d^H\boldsymbol\varsigma\boldsymbol\varsigma^H\boldsymbol
d\boldsymbol d^H\boldsymbol\varsigma]-2\mathbb
E[\boldsymbol\varsigma^H\boldsymbol d\boldsymbol
d^H\boldsymbol\varsigma]+1.
\end{equation}

Since $d_k$ are independent random variables, the evaluation of the
first two terms in the brackets in (\ref{36}) reads
\begin{equation}\label{37}
\begin{split}
&\boldsymbol\varsigma^H\boldsymbol d\boldsymbol
d^H\boldsymbol\varsigma\boldsymbol\varsigma^H\boldsymbol
d\boldsymbol
d^H\boldsymbol\varsigma=\sum_{k=0}^{q-1}\sum_{l=0}^{q-1}|d_k|^2|d_l|^2\varsigma_k^{\ast}\varsigma_k\varsigma_l^{\ast}\varsigma_l,\\
&\boldsymbol\varsigma^H\boldsymbol d\boldsymbol
d^H\boldsymbol\varsigma=\sum_{k=0}^{q-1}|d_k|^2\varsigma_k^{\ast}\varsigma_k^{\ast}.
\end{split}
\end{equation}

For the reduced-rank scheme with the DFP structure, we have
$\boldsymbol w=\boldsymbol T_r\bar{\boldsymbol w}$. Thus,
\begin{equation}\label{38}
\varsigma_k=C_k(\boldsymbol T_r\bar{\boldsymbol w})^H\boldsymbol
a(\theta_k)=C_k\sum_{j=1}^{r}\boldsymbol t_{\bar{w}_j}^H\boldsymbol
a(\theta_k),
\end{equation}
where $\boldsymbol t_{\bar{w}_j}=\bar{w}_j\boldsymbol t_j\in\mathbb
C^{m\times1}$ and $\boldsymbol t_j$ ($j=1, \ldots, r$) is the column
vector of the transformation matrix $\boldsymbol T_r$.

Given $t_j(\theta_k)=C_k\boldsymbol t_{\bar{w}_j}^H\boldsymbol
a(\theta_k)$ and $\varsigma_k=\sum_{j=1}^{r}t_j(\theta_k)$, we get
\begin{equation}\label{39}
\varsigma_k^{\ast}\varsigma_k=\sum_{j=1}^{r}\sum_{n=1}^{r}t_j^{\ast}(\theta_k)t_n(\theta_k).
\end{equation}

From (\ref{38}) and the constraint condition in (\ref{11}), it is
interesting to find $\varsigma_0=C_0\gamma$. Substituting this
expression and (\ref{38}) into (\ref{37}), we have
\begin{equation}\label{40}
\begin{split}
\boldsymbol\varsigma^H\boldsymbol d\boldsymbol
d^H\boldsymbol\varsigma&=|d_0|^2\varsigma_0^{\ast}\varsigma_0+\sum_{k=1}^{q-1}|d_k|^2
\varsigma_k^{\ast}\varsigma_k\\
&=|d_0|^2C_0^2\gamma^2+\tilde{\boldsymbol\varsigma}^H\tilde{\boldsymbol
d}\tilde{\boldsymbol d}^H\tilde{\boldsymbol\varsigma},
\end{split}
\end{equation}
where $\tilde{\boldsymbol\varsigma}=[\varsigma_1, \ldots,
\varsigma_{q-1}]^T\in\mathbb C^{(q-1)\times1}$ and
$\tilde{\boldsymbol d}=[d_1, \ldots, d_{q-1}]^T\in\mathbb
C^{(q-1)\times1}$.

Substituting (\ref{40}) into (\ref{36}), we get the CCM cost
function in the form of reduced-rank, i.e.,
\begin{equation}\label{41}
\begin{split}
J_{\textrm{ccm}}=&\mathbb
E[|d_0|^2C_0^2\gamma^2+\tilde{\boldsymbol\varsigma}^H\tilde{\boldsymbol
d}\tilde{\boldsymbol d}^H\tilde{\boldsymbol\varsigma}]^2\\
&-2\mathbb
E[|d_0|^2C_0\gamma^2+\tilde{\boldsymbol\varsigma}^H\tilde{\boldsymbol
d}\tilde{\boldsymbol d}^H\tilde{\boldsymbol\varsigma}]+1,
\end{split}
\end{equation}
where $\tilde{\boldsymbol\varsigma}$ is a function of the
transformation matrix and the reduced-rank weight vector, as shown
in (\ref{38}). This expression is important for the reduced-rank CCM
analysis. The fact that $\boldsymbol T_r$ and $\bar{\boldsymbol w}$
depend on each other and exchange information claims that we need to
take both of them into consideration for the analysis. The
expression in (\ref{38}) combines these two quantities together and
thus circumvents the complicated procedures of performing the
analysis separately. Note that (\ref{41}) is a constrained
expression since the constraint condition has been enclosed in the
first term of each bracket.

We can examine the convexity of (\ref{11}) by computing the Hessian
matrix $\boldsymbol H$ with respect to
$\tilde{\boldsymbol\varsigma}^H$ and $\tilde{\boldsymbol\varsigma}$,
that is $\boldsymbol
H=\frac{\partial}{\partial\tilde{\boldsymbol\varsigma}^H}\frac{\partial
J_{\textrm{ccm}}}{\partial\tilde{\boldsymbol\varsigma}}$ yields,
\begin{equation}\label{42}
\boldsymbol H=2\mathbb
E\big[(|d_0|^2C_0^2\gamma^2-1)\tilde{\boldsymbol
d}\tilde{\boldsymbol
d}^H+\tilde{\boldsymbol\varsigma}^H\tilde{\boldsymbol
d}\tilde{\boldsymbol
d}^H\tilde{\boldsymbol\varsigma}\tilde{\boldsymbol
d}\tilde{\boldsymbol d}^H+\tilde{\boldsymbol d}\tilde{\boldsymbol
d}^H\tilde{\boldsymbol\varsigma}\tilde{\boldsymbol\varsigma}^H\tilde{\boldsymbol
d}\tilde{\boldsymbol d}^H\big],
\end{equation}
where $\boldsymbol H$ should be positive semi-definite to ensure the
convexity of the optimization problem. The second and third terms in
(\ref{42}) yield positive semi-definite matrices, while the first
term provides the condition $|d_0|^2C_0^2\gamma^2-1\geq0$ to ensure
the convexity. Thus, $J_{\textrm{ccm}}$ is a convex function of
$\boldsymbol T_r$ and $\bar{\boldsymbol w}$ when
\begin{equation}\label{43}
\gamma^2\geq\frac{1}{|d_0|^2C_0^2}.
\end{equation}

For the reduced-rank scheme with the GSC structure, the expression
of the weight vector has been given in (\ref{9}). Substituting this
expression into the definition of $\varsigma_k$ and considering the
fact that $\boldsymbol B\boldsymbol a(\theta_0)=\boldsymbol 0$, we
obtain
\begin{equation}\label{44}
\varsigma_k=\left\{\begin{array}{cc}
                   C_0\boldsymbol a_{\gamma}^H(\theta_0)\boldsymbol
                   a(\theta_0) & \textrm{for}~k=0\\
                   -C_k\sum_{j=1}^{r}\boldsymbol
                   t_{\bar{w}_{\textrm{gsc},j}}^H\boldsymbol
                   B\boldsymbol a(\theta_k) & \textrm{for}~k=1, \ldots,
                   q-1\\
                   \end{array}\right.,
\end{equation}
where $\boldsymbol
t_{\bar{w}_{\textrm{gsc},j}}=\bar{w}_{\textrm{gsc},j}\boldsymbol
t_{\textrm{gsc},j}\in\mathbb C^{(m-1)\times1}$ and $\boldsymbol
t_{\textrm{gsc},j}$ ($j=1, \ldots, r$) is the column vector of
$\boldsymbol T_{\textrm{gsc}}$ for the GSC structure.

Given $t_{j,l}^{'}(\theta_k)=C_k a_l(\theta_k)\boldsymbol
t_{\bar{w}_{\textrm{gsc}},j}^H\boldsymbol b_l$ ($l=1, \ldots, m$),
where $a_n(\theta_k)$ is the $l$th element of the steering vector
with the direction $\theta_k$ and $\boldsymbol b_l\in\mathbb
C^{(m-1)\times1}$ is the $l$th column vector of the signal blocking
matrix $\boldsymbol B$, we have
$\varsigma_k=-\sum_{n=1}^{m}\sum_{j=1}^{r}t_{j,n}^{'}(\theta_k)$.
Thus, for $k=1, \ldots, q-1$,
\begin{equation}\label{45}
\varsigma_k^{\ast}\varsigma_k=\sum_{j=1}^{r}\sum_{n=1}^{r}\sum_{l=1}^{m}\sum_{p=1}^{m}{t_{j,l}^{'}}^{\ast}(\theta_k)t_{n,p}^{'}(\theta_k).
\end{equation}

Substituting (\ref{44}) and (\ref{45}) into (\ref{37}), we get the
expression for the GSC structure, which is
\begin{equation}\label{46}
\begin{split}
\boldsymbol\varsigma^H\boldsymbol d\boldsymbol
d^H\boldsymbol\varsigma&=|d_0|^2\varsigma_0^{\ast}\varsigma_0+\sum_{k=1}^{q-1}|d_k|^2\varsigma_k^{\ast}\varsigma_k\\
&=|d_0|^2C_0^2\boldsymbol a^H(\theta_0)\boldsymbol
a_{\gamma}(\theta_0)\boldsymbol a^H_{\gamma}(\theta_0)\boldsymbol
a(\theta_0)+\tilde{\boldsymbol\varsigma}^H\tilde{\boldsymbol
d}\tilde{\boldsymbol d}^H\tilde{\boldsymbol\varsigma}\\
&=|d_0|^2C_0^2\gamma^2+\tilde{\boldsymbol\varsigma}^H\tilde{\boldsymbol
d}\tilde{\boldsymbol d}^H\tilde{\boldsymbol\varsigma},
\end{split}
\end{equation}
which is in the same form as in (\ref{40}) for the DFP structure but
with the different expression of the quantity
$\tilde{\boldsymbol\varsigma}$. Using the similar interpretation for
the DFP, the quantity $\varsigma_k$ in (\ref{44}) combines the
transformation matrix and the reduced-rank weight vector together
and thus simplifies the analysis. By computing the Hessian matrix
$\boldsymbol H$, we can obtain the same conclusion as shown in
(\ref{43})}.

\section{Simulations}

In this section, we evaluate the output
signal-to-interference-plus-noise ratio (SINR) performance of the
proposed adaptive reduced-rank algorithms and compare them with the
existing methods. Specifically, we compare the proposed SG and RLS
type algorithms with the full-rank (FR) SG and RLS and reduced-rank
methods based on the MSWF and the AVF techniques for both the DFP
and the GSC structures. With respect to each algorithm, we consider
the CMV and the CCM criteria for the beamformer design. We assume
that the DOA of the desired user is known by the receiver. In each
experiment, a total of $K=1000$ runs are carried out to get the
curves. For all simulations, the source power (including the desired
user and interferers) is
$\sigma_{\textrm{s}}^{2}=\sigma_{\textrm{i}}^{2}=1$, the input
signal-to-noise (SNR) ratio is SNR=$10$ dB with spatially and
temporally white Gaussian noise, and $\gamma=1$. Simulations are
performed by an ULA containing $m=32$ sensor elements with
half-wavelength interelement spacing.

\subsection{Comparison of CMV and CCM Based Algorithms}
\begin{figure}[!htb]
\begin{center}
\def\epsfsize#1#2{1.0\columnwidth}
\epsfbox{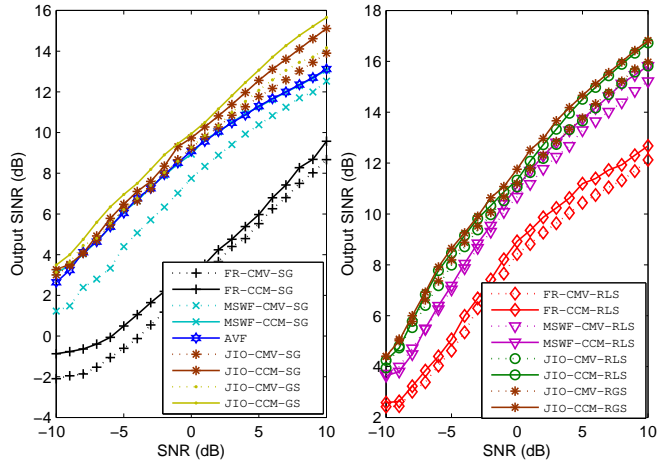} \caption{Output SINR versus input SNR with
$m=32$, $q=5$, SNR$=10$ dB, (a) $\mu_{T_r}=0.002$,
$\mu_{\bar{w}}=0.001$, $r=5$ for SG, $\mu_{T_r}=0.003$,
$\mu_{\bar{w}}=0.0007$, $r=5$ for GS; (b) $\alpha=0.998$,
$\delta=\bar{\delta}=0.03$, $r=5$ for RLS, $\alpha=0.998$,
$\delta=\bar{\delta}=0.028$, $r=5$ for RGS of the proposed CCM
reduced-rank scheme.} \label{fig:cmv_ccm_sinr_gram_final3}
\end{center}
\end{figure}

In this part, we compare the proposed and existing algorithms
according to the CMV and the CCM criteria for the DFP structure of
the beamformer design. The simulation, which includes two
experiments, shows the input SNR versus the output SINR. The input
SNR is varied between $-10$ dB and $10$ dB. The number of users is
$q=5$ with one desired user. Fig. \ref{fig:cmv_ccm_sinr_gram_final3}
(a) plots the curves of the SG type algorithms based on the
full-rank, MSWF, AVF and the proposed reduced-rank scheme, and Fig.
\ref{fig:cmv_ccm_sinr_gram_final3} (b) shows the corresponding RLS
type algorithms. The parameters used to obtain these curves are
given and the rank $r$ is selected to optimize the algorithms. From
Fig. \ref{fig:cmv_ccm_sinr_gram_final3} (a), the output SINR of all
SG type methods increases following the increase of the input SNR.
The algorithms based on the CCM beamformer design outperform those
based on the CMV since the CCM criterion is a positive measure of
the beamformer output deviating from a constant modulus, which
provides more information than the CMV for the parameter estimation
of constant modulus constellations. The proposed CCM algorithms
achieve better performance than the existing full-rank, MSWF and AVF
ones. By employing the GS technique to reformulate the
transformation matrix, the GS version algorithms achieve improved
performance. Fig. \ref{fig:cmv_ccm_sinr_gram_final3} (b) verifies
the same fact but for the RLS type algorithms. It is clear that the
RLS type algorithms superior to the SG type ones for all input SNR
values.

This simulation verifies that the performance of the adaptive
algorithms based on the CCM beamformer design has a similar trend
but is better than that based on the CMV for constant modulus
constellations. Considering this fact, we will only compare the CCM
based adaptive algorithms in the following part for simplification.
Note that all the methods in this simulation are for the DFP
structure. The algorithms for the GSC structure show a similar
performance, which is given in the next part.

\subsection{Output SINR for the DFP and the GSC}
\begin{figure}[!htb]
\begin{center}
\def\epsfsize#1#2{1.0\columnwidth}
\epsfbox{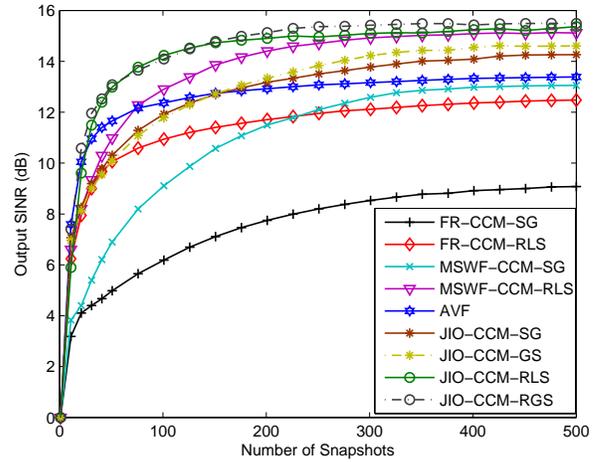} \caption{Output SINR versus the number of
snapshots with $m=32$, $q=7$, SNR$=10$ dB, $\mu_{T_r}=0.003$,
$\mu_{\bar{w}}=0.003$, $r=5$ for SG, $\mu_{T_r}=0.0023$,
$\mu_{\bar{w}}=0.003$, $r=5$ for GS, $\alpha=0.998$,
$\delta=\bar{\delta}=0.025$, $r=5$ for RLS, $\alpha=0.998$,
$\delta=\bar{\delta}=0.02$, $r=5$ for RGS of the DFP structure.}
\label{fig:ccm_allmethods_dsp_gram_final3}
\end{center}
\end{figure}

\begin{figure}[!htb]
\begin{center}
\def\epsfsize#1#2{1.0\columnwidth}
\epsfbox{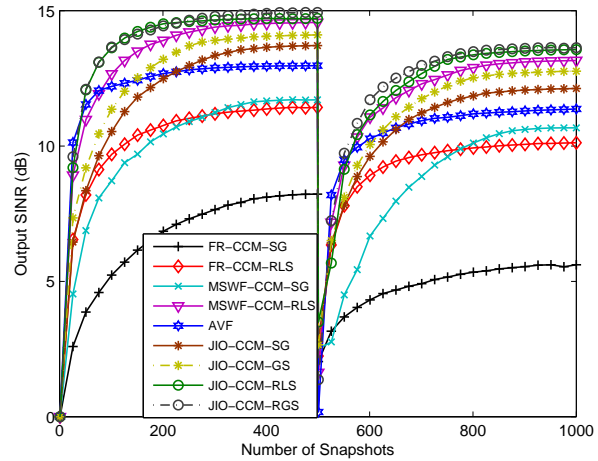} \caption{Output SINR versus input SNR with
$m=32$, $q=7$, SNR$=10$ dB, $\mu_{T_r}=0.0025$,
$\mu_{\bar{w}_{\textrm{gsc}}}=0.002$, $r=5$ for SG,
$\mu_{T_r}=0.003$, $\mu_{\bar{w}_{\textrm{gsc}}}=0.003$, $r=5$ for
GS, $\alpha=0.998$, $\delta=\bar{\delta}=0.01$, $r=5$ for RLS,
$\alpha=0.998$, $\delta=\bar{\delta}=0.0093$, $r=5$ for RGS of the
GSC structure.} \label{fig:ccm_allmethods_gsc_gram_final3}
\end{center}
\end{figure}

We evaluate the output SINR performance of the proposed and existing
algorithms against the number of snapshots for both the DFP and the
GSC structures in Fig. \ref{fig:ccm_allmethods_dsp_gram_final3} and
Fig. \ref{fig:ccm_allmethods_gsc_gram_final3}, respectively. The
number of snapshots is $N=500$. In Fig.
\ref{fig:ccm_allmethods_dsp_gram_final3}, the convergence of the
proposed SG type and extended GS version algorithms is close to the
RLS type algorithm based on the MSWF, and the output SINR values are
higher than other SG type methods based on the full-rank, MSWF and
AVF. The convergence of the proposed RLS type and GS version
algorithms is slightly slower than the AVF, but much faster than
other existing and proposed methods. Its tracking performance
outperforms the MSWF and AVF based algorithms.

Fig. \ref{fig:ccm_allmethods_gsc_gram_final3} is carried out for the
GSC structure under the same scenario as in Fig.
\ref{fig:ccm_allmethods_dsp_gram_final3}. The curves of the
considered algorithms for the GSC show nearly the same convergence
and tracking performance as those for the DFP. It implies that the
GSC structure is an alternative way for the CCM beamformer design.
The difference is that the GSC processor incorporates the constraint
in the structure and thus converts the constrained optimization
problem into an unconstrained one. The adaptive implementation of
the GSC beamformer design is different from that of the DFP but the
performance is similar. The following simulations are carried out
for the DFP structure to simplify the presentation.

\subsection{  Mean Square Estimation Error of the Weight Solution}
\begin{figure}[!htb]
\begin{center}
\def\epsfsize#1#2{1.0\columnwidth}
\epsfbox{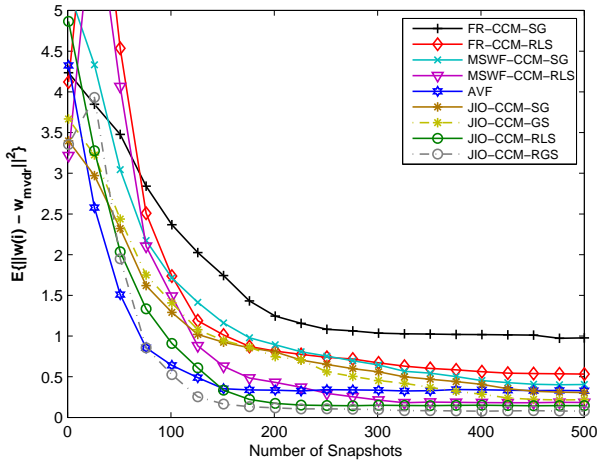} \caption{Square estimation error between the
weight solution and the MVDR weight solution.} \label{fig:weight}
\end{center}
\end{figure}

{  In Fig. \ref{fig:weight}, we measure the mean square estimation
error $\mathbb E\{\|\boldsymbol w(i)-\boldsymbol
w_{\textrm{mvdr}}\|^2\}$ between the weight solutions (full-rank) of
the proposed methods $\boldsymbol w(i)=\boldsymbol
T_r(i)\bar{\boldsymbol w}(i)$ and that of the
minimum-variance-distortionless-response (MVDR) method \cite{Trees}
$\boldsymbol w_{\textrm{mvdr}}=\gamma\frac{\boldsymbol
R^{-1}\boldsymbol a(\theta_0)}{\boldsymbol a^H(\theta_0)\boldsymbol
R^{-1}\boldsymbol a(\theta_0)}$, where $\boldsymbol R$ is obtained
by its sample-average estimation. The experiment is carried out with
the same scenario as in Fig
\ref{fig:ccm_allmethods_dsp_gram_final3}. It exhibits that the mean
square estimation error decreases following the snapshots. The
values of the proposed SG and RLS type algorithms decrease rapidly
and reach a relative lower level compared with those of the existing
methods. Note that $\boldsymbol w_{\textrm{mvdr}}$ is not an optimum
solution for the proposed algorithms but viewed as a referenced
weight solution since, for the CCM based algorithm, the weight
expression is not a pure function of the received data but also
depends on the previous weighting values}.

\subsection{Output SINR versus Rank $r$ and Automatic Rank Selection}
\begin{figure}[!htb]
\begin{center}
\def\epsfsize#1#2{1.0\columnwidth}
\epsfbox{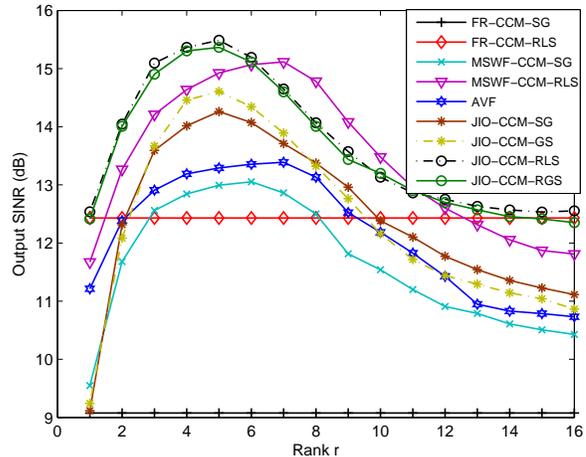} \caption{Output SINR versus rank $r$ with $m=32$,
$q=7$, SNR$=10$ dB.} \label{fig:ccm_rank_gram_final3}
\end{center}
\end{figure}

In the next two experiments, we assess the output SINR performance
of the proposed and analyzed algorithms versus their associated rank
$r$, and check the effectiveness of the automatic rank selection
technique. The experiment in Fig. \ref{fig:ccm_rank_gram_final3} is
intended for setting the adequate rank $r$ of the reduced-rank
schemes for a given input SNR and number of snapshots. The scenario
is the same as that in Fig. \ref{fig:ccm_allmethods_dsp_gram_final3}
except that the number of snapshots is fixed to be $N=500$ and the
rank $r$ is varied between $1$ and $16$. The result indicates that
the best performance of the proposed SG and RLS type algorithms is
obtained with rank $r=5$ for the proposed reduced-rank scheme. The
performance of the full-rank methods is invariant with the change of
the rank $r$. For the MSWF technique, its SG and RLS type algorithms
achieve their best performance with ranks $r=6$ and $r=7$,
respectively. For the AVF-based algorithm, the best rank is found to
be $r=7$. It is interesting to note that the best $r$ is usually
much smaller than the number of elements $m$, which leads to
significant computational savings. For the proposed and analyzed
algorithms, the range of $r$ that has the best performance is
concentrated between $r_{\textrm{min}}=3$ and $r_{\textrm{max}}=7$.
This range is used in the next experiment to check the performance
of the proposed algorithms with the automatic rank selection
technique.

Since the performance of the proposed reduced-rank algorithms was
found in our studies to be a function of the rank $r$ and other
parameters such as the step size and the forgetting factor, we need
to consider their impacts on the performance of the system.
Specifically, we assume that the step size of the SG type algorithms
and the forgetting vector of the RLS type algorithms are adequately
chosen and we focus on the developed automatic rank selection
technique introduced in the previous section.

\begin{figure}[!htb]
\begin{center}
\def\epsfsize#1#2{1.0\columnwidth}
\epsfbox{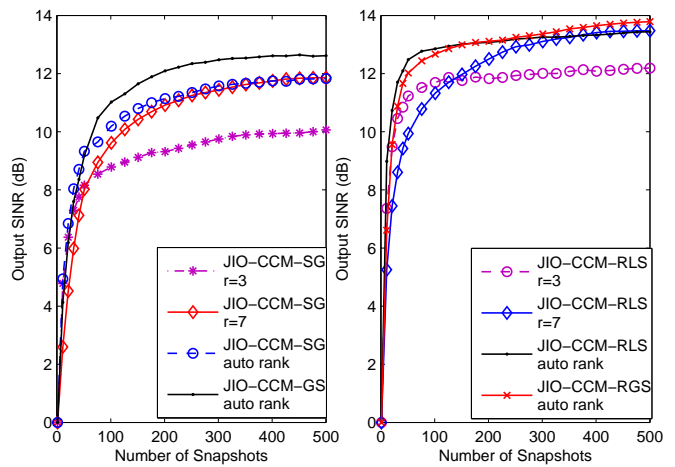} \caption{Output SINR versus the number of
snapshots with $m=32$, $q=10$, SNR$=10$ dB, (a) $\mu_{T_r}=0.003$,
$\mu_{\bar{w}}=0.004$ for SG, $\mu_{T_r}=0.003$,
$\mu_{\bar{w}}=0.001$ for GS; (b) $\alpha=0.998$,
$\delta=\bar{\delta}=0.03$ for RLS, $\alpha=0.998$,
$\delta=\bar{\delta}=0.026$, $r=5$ for RGS with the automatic rank
selection technique.} \label{fig:ccm_rr_sg_rls_autorank_gram_final3}
\end{center}
\end{figure}

In Fig. \ref{fig:ccm_rr_sg_rls_autorank_gram_final3}, the proposed
reduced-rank algorithms utilize fixed values for their rank and also
the automatic rank selection technique. We consider the presence of
$q=10$ users (one desired) in the system. The results show that with
a lower rank $r=3$ the reduced-rank algorithms usually converge
faster but achieve lower output SINR values. Conversely, with a
higher rank $r=7$ the proposed algorithms converge relatively slower
than with a lower rank but reach higher output SINR values. The
developed automatic rank selection technique allows the proposed
algorithms to circumvent the tradeoff between convergence and
steady-state performance for a given rank, by adaptively choosing
the best rank for a given number of snapshots which provides both
fast convergence and improved tracking performance.

\subsection{Performance in non-stationary scenarios}

\begin{figure}[!htb]
\begin{center}
\def\epsfsize#1#2{1.0\columnwidth}
\epsfbox{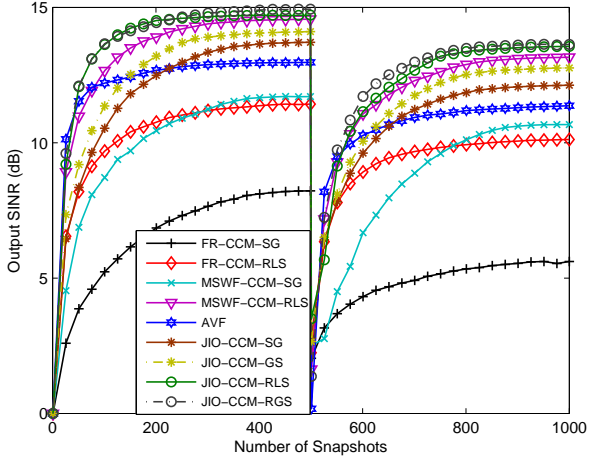} \caption{Output SINR versus input SNR with
$m=32$, $q_{1}=8$, $q_2=11$, SNR$=10$ dB, $\mu_{T_r}=0.003$,
$\mu_{\bar{w}}=0.0038$, $r=5$ for SG, $\mu_{T_r}=0.003$,
$\mu_{\bar{w}}=0.001$, $r=5$ for GS, $\alpha=0.998$,
$\delta=\bar{\delta}=0.033$, $r=5$ for RLS, $\alpha=0.998$,
$\delta=\bar{\delta}=0.028$, $r=5$ for RGS of the proposed CCM
reduced-rank scheme.}
\label{fig:ccm_allmethods_moreusers_gram_final3}
\end{center}
\end{figure}

In the last experiment, we evaluate the performance of the proposed
and analyzed algorithms in a non-stationary scenario, namely, when
the number of users changes. The automatic rank selection technique
is employed, and the step size and the forgetting factor are set to
ensure that the considered algorithms converge quickly to the
steady-state. In this experiment, the scenario starts with $q=8$
users including one desired user. From the first stage (first $500$
snapshots) of Fig. \ref{fig:ccm_allmethods_moreusers_gram_final3},
the convergence and steady-state performance of the proposed SG type
algorithms is superior to other SG type methods with the full-rank,
MSWF and AVF. The proposed RLS type algorithm has a convergence rate
a little slower than the AVF but faster than the other analyzed
methods, and the steady-state performance better than the existing
ones. Three more interferers enter the system at time instant
$i=500$. This change makes the output SINR reduce suddenly and
degrades the performance of all methods. The proposed SG and RLS
type algorithms keep faster convergence and better steady-state
performance in comparison with the corresponding SG and RLS type
methods based on the full-rank and MSWF techniques. The convergence
of the AVF method is fast but the steady-state performance is
inferior to the proposed methods.

\section{Concluding Remarks}
We proposed a CCM reduced-rank scheme based on the joint iterative
optimization of adaptive filters for beamformer design. In the
proposed scheme, the dimension of the received vector is reduced by
the adaptive transformation matrix that is formed by a bank of
full-rank adaptive filters, and the transformed received vector is
processed by the reduced-rank adaptive filter for estimating the
desired signal. The proposed scheme was developed for both DFP and
GSC structures. We derived the CCM expressions for the
transformation matrix and the reduced-rank weight vector, and
developed SG and RLS type algorithms for their efficient
implementation. The GS technique was employed in the proposed
algorithms to reformulate the transformation matrix and thus improve
the performance. The automatic rank selection technique was
developed to determine the most adequate rank and make a good
trade-off between the convergence rate and the steady-state
performance for the proposed methods. The complexity and convexity
analysis of the proposed algorithms was carried out. Simulation
results for the beamforming application showed that the proposed
reduced-rank algorithms significantly outperform the existing
full-rank and reduced-rank methods in convergence and steady-state
performance at comparable complexity.

\begin{appendix}

\section*{Derivation of (\ref{24})}
In this appendix, we show the details of the derivation of the
expression for the transformation matrix in (\ref{33}). Assuming
$\bar{\boldsymbol w}(i)\neq\boldsymbol 0$ is known, taking the
gradient terms of (\ref{32}) with respect to $\boldsymbol T_r(i)$,
we get
\begin{equation}\label{62}
\begin{split}
\nabla L_{\textrm{un}_{\boldsymbol
T_r(i)}}&=2\sum_{l=1}^{i}|y(l)|^2\boldsymbol x(l)\boldsymbol
x^H(l)\boldsymbol T_r(i)\bar{\boldsymbol w}(i)\bar{\boldsymbol
w}^H(i)\\
&-2\sum_{l=1}^{i}\boldsymbol x(l)\boldsymbol x^H(l)\boldsymbol
T_r(i)\bar{\boldsymbol w}(i)\bar{\boldsymbol
w}^H(i)+2\lambda\boldsymbol a(\theta_0)\bar{\boldsymbol w}^H(i)\\
&=2\hat{\boldsymbol R}(i)\boldsymbol T_r(i)\bar{\boldsymbol
w}(i)\bar{\boldsymbol w}^H(i)-2\hat{\boldsymbol
p}(i)\bar{\boldsymbol w}^H(i)\\
&+2\lambda\boldsymbol a(\theta_0)\bar{\boldsymbol w}^H(i).
\end{split}
\end{equation}

Making the above gradient terms equal to the zero matrix,
right-multiplying the both sides by $\bar{\boldsymbol w}(i)$, and
rearranging the expression, it becomes
\begin{equation}\label{63}
\boldsymbol T_r(i)\bar{\boldsymbol w}(i)=\hat{\boldsymbol
R}^{-1}(i)\big[\hat{\boldsymbol p}(i)-\lambda\boldsymbol
a(\theta_0)\big]
\end{equation}

If we define $\hat{\boldsymbol p}_{\hat{R}}(i)=\hat{\boldsymbol
R}^{-1}(i)\big[\hat{\boldsymbol p}(i)-\lambda\boldsymbol
a(\theta_0)\big]$, the solution of $\boldsymbol T_r(i)$ in
(\ref{63}) can be regarded to find the solution to the linear
equation
\begin{equation}\label{64}
\boldsymbol T_r(i)\bar{\boldsymbol w}(i)=\hat{\boldsymbol
p}_{\hat{R}}(i)
\end{equation}

Given a $\bar{\boldsymbol w}(i)\neq\boldsymbol 0$, there exists
multiple $\boldsymbol T_r(i)$ satisfying (\ref{64}) in general.
Therefore, we derive the minimum Frobenius-norm solution for
stability. Let us express the quantities involved in (\ref{64}) by
\begin{equation}\label{65}
\boldsymbol T_{r}(i)=\begin{bmatrix}
 \bar{\boldsymbol\rho}_{1}(i)\\
 \bar{\boldsymbol\rho}_{2}(i)\\
 \vdots\\
 \bar{\boldsymbol\rho}_{m}(i)\\ \end{bmatrix};~~
%\bar{\boldsymbol w}(i)=\begin{bmatrix}
% \bar{w}_1(i)\\
% \bar{w}_2(i)\\
% \vdots\\
% \bar{w}_r(i)\\ \end{bmatrix};~~
\hat{\boldsymbol p}_{\hat{R}}(i)=\begin{bmatrix}
 \hat{p}_{\hat{R},1}(i)\\
 \hat{p}_{\hat{R},2}(i)\\
 \vdots\\
 \hat{p}_{\hat{R},m}(i)\\ \end{bmatrix}
\end{equation}

The search for the minimum Frobenius-norm solution of (\ref{64}) is
reduced to the following $m$ subproblems $(j=1, \ldots, m)$:
\begin{equation}\label{66}
\min\|\bar{\boldsymbol\rho}_{j}(i)\|^2~~~ \textrm{subject~to}~~
\bar{\boldsymbol\rho}_{j}(i)\bar{\boldsymbol
w}(i)=\hat{p}_{\hat{R},j}(i),
\end{equation}

The solution to (\ref{66}) is the projection of
$\bar{\boldsymbol\rho}(i)$ onto the hyperplane $\mathcal
{H}_{j}(i)=\big\{\bar{\boldsymbol \rho}(i)\in\mathbb C^{1\times
r}:~\bar{\boldsymbol\rho}(i)\bar{\boldsymbol
w}(i)=\hat{p}_{\hat{R},j}(i)\big\}$, which is given by
\begin{equation}\label{67}
\bar{\boldsymbol\rho}_{j}(i)=\hat{p}_{\hat{R},j}(i)\frac{\bar{\boldsymbol
w}^H(i)}{\|\bar{\boldsymbol w}(i)\|^2}
\end{equation}

Hence, the minimum Frobenius-norm solution of the transformation
matrix is given by
\begin{equation}\label{68}
\boldsymbol T_r(i)=\hat{\boldsymbol
p}_{\hat{R}}(i)\frac{\bar{\boldsymbol w}^H(i)}{\|\bar{\boldsymbol
w}(i)\|^2}
\end{equation}

Substituting the definition of $\hat{\boldsymbol p}_{\hat{R}}(i)$
into (\ref{68}), we have
\begin{equation}\label{69}
\boldsymbol T_r(i)=\hat{\boldsymbol R}^{-1}(i)\big[\hat{\boldsymbol
p}(i)-\lambda\boldsymbol a(\theta_0)\big]\frac{\bar{\boldsymbol
w}^H(i)}{\|\bar{\boldsymbol w}(i)\|^2}
\end{equation}

The multiplier $\lambda$ can be obtained by incorporating (\ref{63})
with the constraint $\bar{\boldsymbol w}^H(i)\boldsymbol
T_r^H(i)\boldsymbol a(\theta_0)=\gamma$, which is
\begin{equation}\label{70}
\lambda=\frac{\hat{\boldsymbol p}(i)\hat{\boldsymbol
R}^{-1}(i)\boldsymbol a(\theta_0)-\gamma}{\boldsymbol
a^H(\theta_0)\hat{\boldsymbol R}^{-1}(i)\boldsymbol a(\theta_0)}
\end{equation}

Therefore, the expression of the transformation matrix in (\ref{33})
can be obtained by substituting (\ref{70}) into (\ref{69}).

\end{appendix}

\end{document}